\documentclass[aps,prd,reprint,preprintnumbers,showpacs,nofootinbib,superscriptaddress,longbibliography]{revtex4-2}
\usepackage{natbib}

\usepackage[utf8]{inputenc}
\usepackage[T1]{fontenc}
\usepackage[english]{babel}

\usepackage{amsmath,amsfonts,amssymb}
\usepackage{siunitx}
\DeclareSIUnit{\year}{yr}
\DeclareSIUnit{\km}{km}
\DeclareSIUnit{\parsec}{pc}
\usepackage{mathtools}
\usepackage{mathrsfs}
\usepackage{tensor}
\usepackage{accents}
\allowdisplaybreaks

\usepackage{graphicx}
\usepackage[dvipsnames]{xcolor}
\usepackage{tikz}
\usetikzlibrary{shapes.geometric,arrows.meta,positioning}
\usepackage{tikz-feynman}
\tikzfeynmanset{compat=1.1.0}
\makeatletter
\tikzfeynmanset{
  /tikzfeynman/every boson@@/.style={
    double, double distance=1.2pt, line width=0.5pt,
    decoration={snake, amplitude=0.6mm, segment length=3.0mm}, decorate
  }
}
\makeatother

\usepackage{etoolbox}
\usepackage{xspace}
\usepackage{scalerel,stackengine}

\setcitestyle{numbers,comma,square}

\usepackage{hyperref}
\hypersetup{colorlinks=true, linkcolor=Blue, citecolor=Blue, filecolor=Blue, urlcolor=Blue}
\usepackage[capitalize,nameinlink]{cleveref}
\crefname{equation}{Eq.}{Eqs.}
\Crefname{equation}{Eq.}{Eqs.}
\crefname{figure}{Fig.}{Figs.}
\Crefname{figure}{Fig.}{Figs.}
\crefname{section}{Sec.}{Secs.}
\Crefname{section}{Sec.}{Secs.}
\creflabelformat{equation}{#2(#1)#3}
\AddToHook{cmd/appendix/before}{\crefalias{section}{appendix}}

\makeatletter
\renewcommand{\paragraph}{%
\@startsection{paragraph}{4}%
{\z@}{1.21ex \@plus 1ex \@minus .2ex}{0.9em}%
{\normalfont\normalsize\bfseries}%
}
\usepackage{titlesec}
\newrobustcmd{\pea}[1]{\emph{#1}\textbf{.\ \ \ ---}}
\titleformat{\paragraph}[runin]{\normalfont\normalsize\bfseries}{\emph\theparagraph}{1em}{\pea}
\titleformat{\section}[block]{\normalfont\bfseries\centering}{\MakeUppercase\thesection}{1em}{\MakeUppercase}
\makeatother

\makeatletter \def\switch@array{}\makeatother

\newrobustcmd{\TensorField}[1]{%
	\tensor{\mathcal{H}}{#1}%
}
\newrobustcmd{\VectorField}[1]{%
	\tensor{\mathcal{A}}{#1}%
}
\newrobustcmd{\ScalarField}{%
	\phi%
}
\newrobustcmd{\VectorGauge}[1]{%
	\tensor{\xi}{#1}%
}
\newrobustcmd{\ScalarGauge}{%
	\chi%
}
\newrobustcmd{\SymRankThreeField}[1]{%
	\tensor{\mathcal{Q}}{#1}%
}
\newrobustcmd{\AntRankThreeField}[1]{%
	\tensor{\mathcal{T}}{#1}%
}
\newrobustcmd{\PD}[1]{%
	\tensor{\partial}{#1}%
}
\newrobustcmd{\Coupling}[1]{%
	\tensor{\theta}{#1}%
}
\newrobustcmd{\GeneralCoupling}[2][2]{%
	\tensor*{\theta}{^{(#1)}#2}%
}
\newrobustcmd{\FPCoupling}[1]{%
	\tensor{\bar{\theta}}{#1}%
}
\newrobustcmd{\LamThree}{\Lambda_{3}}
\newrobustcmd{\WMat}[2]{%
	\tensor*{\mathsf{W}}{^{(#1)}_{#2}}%
}
\newrobustcmd{\DeltaOrder}[1]{%
	\delta^{(#1)}%
}
\newrobustcmd{\MassDimension}[1]{%
	\tensor{d}{#1}%
}
\newrobustcmd{\Operator}[1]{%
	\mathcal{O}^{#1}%
}
\newrobustcmd{\MAGG}[2]{%
	\tensor{\vphantom{l^{l^2}}\smash{\stackrel{\raisebox{-3pt}{\scalebox{0.55}{$({#1})$}}}{\scalebox{0.99}{$\theta$}}}}{#2}%
}

\newrobustcmd{\MoorePenroseJP}[2]{%
	\tensor*{\mathsf{O}}{^+_{{#1^#2}}}(\theta ; k)%
}
\newrobustcmd{\WaveOperatorJP}[2]{%
	\tensor*{\mathsf{O}}{_{{#1^#2}}}(\theta ; k)%
}
\newrobustcmd{\WaveOperatorJPk}[3]{%
	\tensor*{\mathsf{O}}{_{{#1^#2}}}(\theta ; #3)%
}
\newrobustcmd{\DetCoeff}[3]{%
	\tensor*{c}{^{{#2^#3}}_{#1}}(\theta)%
}
\newrobustcmd{\SoundHorizonDrag}{r_{\text{d}}}
\newrobustcmd{\WDe}{w_{\text{de}}}
\newrobustcmd{\PDe}{P_{\text{de}}}
\newrobustcmd{\RhoDe}{\rho_{\text{de}}}
\newrobustcmd{\RhoDeDot}{\dot{\rho}_{\text{de}}}
\newrobustcmd{\FDe}{f_{\text{de}}}
\newrobustcmd{\OmegaM}{\Omega_{\text{m}}}
\newrobustcmd{\OmegaMFid}{\Omega_{\text{m}}^{\text{fid}}}
\newrobustcmd{\HZeroFid}{H_0^{\text{fid}}}
\newrobustcmd{\Cov}{\mathsf{C}}
\newrobustcmd{\CovProj}{\widetilde{\mathsf{C}}}
\newrobustcmd{\CovBAO}{\mathsf{C}_{\text{BAO}}}
\newrobustcmd{\CovSNe}{\mathsf{C}_{\text{SNe}}}
\newrobustcmd{\CovSNeProj}{\widetilde{\mathsf{C}}_{\text{SNe}}}
\newrobustcmd{\OneVec}{\mathsf{n}}
\newrobustcmd{\mB}{m_{\text{b}}}
\newrobustcmd{\Lik}{\mathcal{L}}
\newrobustcmd{\LikBAO}{\mathcal{L}_{\text{BAO}}}
\newrobustcmd{\LikSNe}{\mathcal{L}_{\text{SNe}}}
\newrobustcmd{\LikMar}{\mathcal{L}_{\text{mar}}}
\newrobustcmd{\LikMarLCDM}{\mathcal{L}_{\text{mar}}^{\Lambda\text{CDM}}}
\newrobustcmd{\PSurv}{p_{\text{surv}}}
\newrobustcmd{\tBH}{t_{\text{BH}}}
\newrobustcmd{\NFin}{N_{\text{fin}}}
\newrobustcmd{\GammaSR}{\Gamma_{\text{SR}}}
\newrobustcmd{\MBH}{M_{\text{BH}}}
\newrobustcmd{\MPl}{M_{\text{Pl}}}
\newrobustcmd{\aStar}{a_*}
\newrobustcmd{\BHSR}{\emph{BHSR}\xspace}
\newrobustcmd{\mL}{m_l}
\newrobustcmd{\kNN}{k_{\text{NN}}}
\newrobustcmd{\VanZ}[1]{\tensor{z}{_{#1}}}
\newrobustcmd{\VanK}[1]{\tensor{k}{_{#1}}}
\newrobustcmd{\VanV}[2]{\tensor{V}{_{#1#2}}}
\newrobustcmd{\WaveOperatorJPn}[3]{%
	\tensor*{\mathsf{O}}{^{(#3)}_{{#1^#2}}}(\theta)%
}
\newrobustcmd{\NullVector}[3]{%
	\tensor*{\mathsf{v}}{_{#3_{{#1^#2}}}}(\theta ; k)%
}
\newrobustcmd{\NullVectorSoft}[3]{%
	\tensor*{\mathsf{v}}{^{\text{soft}}_{#3_{{#1^#2}}}}(\theta ; k)%
}
\newrobustcmd{\NullVectorm}[4]{%
	\tensor*{\mathsf{v}}{_{#3_{{#1^#2}}}}(\theta ; \tensor*{m}{_{#4_{{#1^#2}}}}(\theta))%
}
\newrobustcmd{\NullVectorN}[4]{%
	\tensor*{\mathsf{v}}{^{(#4)}_{#3_{{#1^#2}}}}(\theta)%
}
\newrobustcmd{\NullVectorConj}[3]{%
	\tensor*{\mathsf{v}}{^\dagger_{#3_{{#1^#2}}}}(\theta ; k)%
}
\newrobustcmd{\NewRes}[2]{%
	\mathop{\mathrm{Res}}_{#1\to#2}%
}
\newrobustcmd{\RegWaveOperatorJP}[2]{%
	\tensor*{\mathsf{O}}{^{\text{reg}}_{{#1^#2}}}(\theta ; k)%
}
\newrobustcmd{\RegWaveOperatorJPms}[3]{%
	\tensor*{\mathsf{O}}{^{\text{reg}}_{{#1^#2}}}(\theta ; \tensor*{m}{_{#3_{{#1^#2}}}}(\theta))%
}
\newrobustcmd{\PhysNullVector}[3]{%
	\tensor*{\mathsf{u}}{_{#3_{{#1^#2}}}}(\theta)%
}
\newrobustcmd{\ProjNullVector}[3]{%
	\tensor*{\mathsf{w}}{_{#3_{{#1^#2}}}}(\theta)%
}
\newrobustcmd{\ProjNullVectorConj}[3]{%
	\tensor*{\mathsf{w}}{^{\dagger}_{#3_{{#1^#2}}}}(\theta)%
}
\newrobustcmd{\GaugeMatrixJP}[2]{%
	\tensor*{\mathsf{V}}{_{{#1^#2}}}(\theta)%
}
\newrobustcmd{\GaugeMatrixJPConj}[2]{%
	\tensor*{\mathsf{V}}{^{\dagger}_{{#1^#2}}}(\theta)%
}
\newrobustcmd{\GaugeProjector}[2]{%
	\tensor*{\mathsf{P}}{_{{#1^#2}}}(\theta)%
}
\newrobustcmd{\WaveOperatorJPprime}[2]{%
	\tensor*{\mathsf{O}}{^{\prime}_{{#1^#2}}}(m(\theta))%
}
\newrobustcmd{\Residue}[3]{%
	\tensor*{Z}{_{#3_{{#1^#2}}}}(\theta)%
}
\newrobustcmd{\Relevance}[3]{%
	\tensor*{R}{_{#3_{{#1^#2}}}}(\theta)%
}
\newrobustcmd{\Likelihood}{%
	\mathcal{L}(\text{QFT}|\theta)%
}
\newrobustcmd{\Mass}[3]{%
	\tensor*{m}{_{#3_{{#1^#2}}}}(\theta)%
}
\newrobustcmd{\BlockMatrix}[2]{%
	\tensor*{\bar{\mathsf{O}}}{_{{#1^#2}}}(\theta)%
}
\newrobustcmd{\BlockNullVector}[3]{%
	\tensor*{\bar{\mathsf{v}}}{_{#3_{{#1^#2}}}}%
}
\newrobustcmd{\BlockLSV}[3]{%
	\tensor*{\bar{\mathsf{u}}}{_{#3_{{#1^#2}}}}%
}
\newrobustcmd{\BlockLSVConj}[3]{%
	\tensor*{\bar{\mathsf{u}}}{^{\dagger}_{#3_{{#1^#2}}}}%
}
\newrobustcmd{\BlockSV}[3]{%
	\tensor*{\sigma}{_{#3_{{#1^#2}}}}(\theta)%
}

\def\a{\alpha}\def\b{\beta}\def\g{\gamma}\def\d{\delta}
\def\e{\epsilon}
\let\polishl\l\DeclareUnicodeCharacter{0142}{\polishl}\def\l{\lambda}
\let\PolishL\L\DeclareUnicodeCharacter{0141}{\PolishL}\def\L{\Lambda}

\def\fr{\frac}\def\tr{{\rm tr}}

\newcommand{\be}{\begin{equation}}
\newcommand{\ee}{\end{equation}}
\newcommand{\bea}{\begin{equation}\begin{aligned}}
\newcommand{\eea}{\end{aligned}\end{equation}}
\newcommand{\WolframLanguage}{\textit{Wolfram Language}\xspace}
\newcommand{\Python}{\textit{Python}\xspace}
\newcommand{\JAX}{\textit{JAX}\xspace}
\newcommand{\BlackJAX}{\textit{BlackJAX}\xspace}
\newcommand{\Optimistix}{\textit{Optimistix}\xspace}

\newcommand{\SOFTSUSY}{\textit{SOFTSUSY}\xspace}
\newcommand{\SARAH}{\textit{SARAH}\xspace}
\newcommand{\SPheno}{\textit{SPheno}\xspace}

\newcommand{\anesthetic}{\textit{anesthetic}\xspace}

\usepackage{soul}

\begin{document}

\title{Numerical polology: towards next-generation model-building for cosmology}

\author{Will Barker}
\email{barker@fzu.cz}
\affiliation{Institute of Physics of the Czech Academy of Sciences, Na Slovance 1999/2, 182 00 Prague 8, Czechia}

\author{Will Handley}
\email{wh260@cam.ac.uk}
\affiliation{Kavli Institute for Cosmology, Madingley Road, Cambridge, CB3 0HA, UK}
\affiliation{Institute of Astronomy, University of Cambridge, Madingley Road, Cambridge, CB3 0HA, UK}

\author{Michael Hobson}
\email{mph@mrao.cam.ac.uk}
\affiliation{Astrophysics Group, Cavendish Laboratory, J.J. Thomson Avenue, Cambridge, CB3 0HE, UK}

\author{Anthony Lasenby}
\email{a.n.lasenby@mrao.cam.ac.uk}
\affiliation{Astrophysics Group, Cavendish Laboratory, J.J. Thomson Avenue, Cambridge, CB3 0HE, UK}
\affiliation{Kavli Institute for Cosmology, Madingley Road, Cambridge, CB3 0HA, UK}

\author{Carlo Marzo}
\email{carlo.marzo@kbfi.ee}
\affiliation{Laboratory for High Energy and Computational Physics, NICPB, R\"{a}vala 10, Tallinn 10143, Estonia}

\author{Alessandro Santoni}
\email{asantoni@uc.cl}
\affiliation{Institute of Space Sciences and Astronomy, University of Malta, Msida, Malta}
\affiliation{Department of Physics, University of Malta, Msida, Malta}

\author{Leonardo Torcellini}
\email{leonardo.torcellini@unisalento.it}
\affiliation{Dipartimento di Matematica e Fisica, Universit\`{a} del Salento and INFN, Sezione di Lecce, Via Arnesano 73100 Lecce, Italy}

\begin{abstract}
	The dark sector need not be restricted to simple field content. Indeed, simple bosonic configurations, such as scalar-tensor or dark photon models, contrast with the much richer picture painted by many ultraviolet scenarios. Polology is the study of propagator poles, which correspond to particle states in any given theory. We outline a numerical polology framework for discovering perturbative, ghost-free models with consistent interactions, which produces theoretical model priors by sampling the coupling space. The method is tested on tensor field theories of up to rank three. Subsequent observational constraint pipelines are illustrated for black hole superradiance (M33~X-7), dynamical dark energy (DESI~DR2, Pantheon and SH0ES) and gravitational waves (GWTC-3).
\end{abstract}

\maketitle

\section{Introduction} \label{intro}

\paragraph*{Theories are data models} Particle dark matter, along with many ultraviolet scenarios, suggest that additional low-energy degrees of freedom remain to be discovered. In a popular scenario, bosonic fields on flat spacetime will form a (generally non-renormalizable) effective field theory (EFT) with action
\be\label{EFTLag}
\mathcal{S}(\Coupling{}) = \int \mathrm{d}^4x \, \sum_i \Coupling{_i} \, \Operator{i},
\ee
in which operators with mass dimension~$[\Operator{i}] = n$ are parameterised by couplings~$\Coupling{_i}\sim\L^{4-n}$, where~$\L$ is a cutoff.

From the theoretician's perspective,~$\mathcal{S}(\Coupling{})$ is a \emph{functional} of the field content; yet it may also be viewed as a \emph{function} of the couplings~$\Coupling{}$, and it is this latter dependence which we choose to emphasise. Theories of physics are inherently subordinate to the observed phenomena: any action~$\mathcal{S}(\Coupling{})$ is ultimately a data model, for which the couplings~$\Coupling{}$ are model parameters. Given observational data~$D$, the relative worth of a theory is then determined by the evidence
\be\label{eq:evidence}
\mathcal{Z}(D) = \int \mathrm{d}\Coupling{}\, \mathcal{L}(D|\Coupling{})\, \pi(\Coupling{}),
\ee
where~$\mathcal{L}(D|\Coupling{})$ is the likelihood of~$D$ in the context of~$\mathcal{S}(\Coupling{})$, and~$\pi(\Coupling{})$ is the prior probability of the~$\Coupling{}$. In precision cosmology, great advances have been made in collecting~$D$ and efficiently computing~$\mathcal{L}(D|\Coupling{})$ for the available~$\mathcal{S}(\Coupling{})$ that seem best-motivated; such fits are systematised in global-fitting frameworks such as \emph{GAMBIT}~\cite{GAMBIT:2021rlp,Balazs:2022tjl,GAMBITCosmologyWorkgroup:2020rmf}, which evaluate~$\mathcal{L}(D|\Coupling{})$ --- and, in a Bayesian analysis, the evidence in~\cref{eq:evidence} itself --- for a given~$\mathcal{S}(\Coupling{})$. The complementary theoretical programme for producing candidate~$\mathcal{S}(\Coupling{})$ is also very active; in relative terms, though, it is less systematic (note counterexamples such as~\cite{Arkani-Hamed:2015bza,Chen:2009zp},~\cite{Cheung:2007st,Weinberg:2008hq} and~\cite{Creminelli:2008wc,Gubitosi:2012hu,Gleyzes:2013ooa}), while the need for efficient computation of~$\pi(\Coupling{})$ is often overlooked.

The ability to compute~$\mathcal{L}(D|\Coupling{})$ for some~$\mathcal{S}(\Coupling{})$ is called \emph{predictivity}. Regardless of renormalizability, a theory of the form given in~\cref{EFTLag} is endowed with predictivity by the systematics of quantum field theory, and by nothing else~\cite{Weinberg:1978kz,Coleman:1969sm,Callan:1969sn,Burgess:2007pt}. Quantum field theory also imposes strict constraints on the permissible values of the~$\Coupling{}$, and it is helpful to interpret these constraints as contributing to~$\pi(\Coupling{})$. For example, the~$\Operator{i}$ which are quadratic in the fields determine the particle spectrum: the unitarity of the~$S$-matrix typically excludes portions of the corresponding~$\Coupling{}$-space. For the free spectrum, this is the requirement of positive propagator residues (no ghosts) and real pole masses (no tachyons); it is distinct from, though complementary to, the energy-dependent perturbative-unitarity bounds that such global fits impose on the interacting couplings~\cite{Chang:2022jgo,Chang:2023cki}.

Nor is this `quantum' contribution to~$\pi(\Coupling{})$ ever \emph{too} restrictive: it avoids contrived models whose~$\Coupling{}$ are subject to ad hoc tunings, i.e., conditionings of the prior which cannot be justified without knowledge of the ultraviolet physics. Specifically, it is the symmetries of the quadratic sector which dictate all the non-linear operators according to the perturbative Noether procedure~\cite{Berends:1984rq,Francia:2016weg}. It is thus not permitted to shrink~$\pi(\Coupling{})$ via any tunings of the~$\Coupling{}$ which eliminate symmetric interactions; doing so would jeopardise the radiative stability of the model~\cite{Marzo:2021iok,Marzo:2024pyn}. Rather, the~$\pi(\Coupling{})$ for a symmetric model may actually be \emph{expanded} into the space of non-symmetric theories, but this process is also tightly controlled. The organising principle is that small, symmetry-breaking~$\Coupling{}$ inherit some radiative protection, and are technically natural~\cite{tHooft:1979rat}. Valid~$\pi(\Coupling{})$ must be derived from all these quantum considerations, and many others besides.

\paragraph*{Scalable model-building} The class of theories in~\cref{EFTLag} successfully describes all the fundamental bosons that are currently known to exist, and thus offers a conservative infrared foundation for new physics. The coordinate invariance of linearised gravity --- as a symmetry --- gives rise uniquely to non-linear general relativity~\cite{Deser:1987uk} (see also~\cite{Butcher:2009ta}) and its predictive EFT extension~\cite{Donoghue:1994dn,Donoghue:2017ovt}; the Maxwell theory meanwhile gives rise to Yang--Mills. Small symmetry-breaking masses for both the vector gauge boson and graviton are technically natural; the former are actually observed in the weak sector, where the symmetry is restored by the Higgs mechanism.

Many proposals for new physics in cosmology concern spin-zero species, but there is also a literature on spin-one~\cite{Nelson:2011sf,Graham:2015rva,Caputo:2021eaa,Pani:2012vp,Pani:2012bp,Baryakhtar:2017ngi,Cardoso:2018tly}, spin-two~\cite{Brito:2013wya,Brito:2020lup} and higher-spin~\cite{Alexander:2020gmv,Falkowski:2020fsu,Jain:2021pnk} particles. For these spins respectively, the Klein--Gordon~\cite{Klein:1926tv,Gordon:1926emj}, Proca~\cite{Proca:1936fbw}, Fierz--Pauli~\cite{Fierz:1939zz,Fierz:1939ix} and Singh--Hagen~\cite{Singh:1974qz} actions are normally used as a foundation, from which interactions are attempted to be derived.\footnote{Of course, interactions from Klein--Gordon and Singh--Hagen theories are respectively nil and problematic, see e.g.~\cite{deWit:1979sib,Klishevich:1997pd,Zinoviev:2008ck}.} For comprehensive model comparison, however, it is important to understand the full range of possibilities, and in particular we observe that:
\begin{center}
	``\emph{Not all theories can be described just by specifying the particle spin; the field content of the action leads to its own phenomenological implications.}''
\end{center}

For example, a symmetric, rank-two tensor field can propagate a spin-two particle, but it can also propagate a scalar if Klein--Gordon theory is written in terms of the trace. This does not affect the phenomenology, but when \emph{multiple} fields are used, the number of possible interaction operators and constraining symmetries grows rapidly. Interactions (i.e. non-linearities) are absolutely crucial to phenomenological signatures. The fact that the quantum~$\pi(\Coupling{})$ may be hard to compute in such cases does not automatically exclude them from participating in the dark sector as technically natural EFTs, and we are in principle obliged to consider them.

The proliferation of the~$\Coupling{}$ due to extended field content may seem undesirable, since extra parameters are sometimes assumed to weaken predictivity through an associated Occam penalty~\cite{Hergt:2021qlh}. In practice, however, this question may only be decided by a detailed application of~\cref{eq:evidence}, whose intimate dependence on~$\pi(\Coupling{})$ just reinforces the importance of correctly computing the prior. Moreover, within the bounds set by~$\pi(\Coupling{})$, it is not unreasonable to expect that high~$\mathcal{L}(D|\Coupling{})$ and competitive~$\mathcal{Z}(D)$ may \emph{only} be achievable in the context of some richly parametric~$\mathcal{S}(\Coupling{})$, whose specific interactions can be used to explain some equally specific signal in~$D$.

In any case, the volume and quality of~$D$ will continue to grow with each generation of surveys. It is thus important that our capacity for the systematic manufacture of~$\mathcal{S}(\Coupling{})$ scales appropriately, and this calls for a high degree of automation.

\paragraph*{In this work} A framework is developed for efficiently computing the unitary prior~$\pi(\Coupling{})$ on the couplings~$\Coupling{}$ which parameterise the free sector of~\cref{EFTLag} --- i.e., the operators~$\Operator{i}$ which are quadratic in the fields. A numerical approach is also available for quantum corrections: these are necessary for constraining interaction couplings under the same framework, which will be the subject of the companion paper. Even at tree level, however, there is scope for connecting~$\mathcal{S}(\Coupling{})$ to cosmological observables: some basic observational constraints on a new Stueckelberg extension of massive gravity are illustrated.

Our technique --- \emph{numerical polology} --- fully embraces the interpretation of~$\mathcal{S}(\Coupling{})$ as a data model, by promoting~$\Coupling{}$-space to the central object of study.

This kind of approach has yet to be established in cosmology, but it has a clear precedent in particle physics through the \SARAH~\cite{Allanach:2001kg}, \SOFTSUSY~\cite{Staub:2008uz,Staub:2013tta} and \SPheno~\cite{Porod:2003um,Porod:2011nf} infrastructures. There are, however, important differences in our approach. Firstly, the method admits non-renormalisable theories, with arbitrarily many higher-rank fields. Secondly, the sampling procedure is incorporated into the algorithm, which allows extensions beyond \emph{model testing} and into \emph{model discovery}.

The remainder of this paper is organised as follows. \cref{overview} provides illustrative examples of the method. \cref{implementation} develops the theory underlying the algorithm, and provides details of the computational implementation. Conclusions follow in~\cref{discussion}, and there are several technical appendices. We use natural units~$\hbar\equiv c\equiv 1$ throughout, and the particle physics signature~$(+,-,-,-)$.

\section{Examples} \label{overview}

\paragraph*{Numerical polology} The particle spectrum of a theory is determined by the poles of the propagator, the study of which is called \emph{polology}~\cite{Weinberg:1995mt}.

The formal polology algorithm for computing the free particle spectrum and associated unitarity of~\cref{EFTLag} has been known since the work of Van Nieuwenhuizen and Sezgin~\cite{VanNieuwenhuizen:1973fi,Neville:1978bk,Neville:1979rb,Sezgin:1981xs,Sezgin:1979zf,Kuhfuss:1986rb,Karananas:2014pxa,Karananas:2016ltn,Mendonca:2019gco,Percacci:2020ddy,Marzo:2021esg,Marzo:2021iok,Mikura:2023ruz,Mikura:2024mji}, but the technique is computationally expensive. Computer algebra implementations such as \emph{PSALTer}~\cite{Barker:2024juc,Barker:2025qmw}, \emph{Kummitus}~\cite{Marzo:2026yjg} and others~\cite{Lin:2018awc,Lin:2019ugq,Lin:2020phk,Lin:2020fuo} have recently been developed, but these scale poorly with the complexity of the theory due to expression swell. For a scalable approach, the only avenue is numerical.

In this section the scope of the method is illustrated with a series of examples. \cref{tuned} considers simple models, before the approach is used as a tool for theoretical discovery in~\cref{untuned}. \cref{results} extends beyond theory to show that the framework natively accommodates data-driven constraints, and interactions are introduced in~\cref{interactions_main}.

\subsection{Tuned theories} \label{tuned}

\paragraph*{Fierz--Pauli theory} Linear gravity is expressed in terms of the perturbation of the metric around the Minkowski vacuum, which can be represented by a symmetric rank-two field~$\TensorField{_\a_\b} \equiv \TensorField{_{(\a\b)}}$ with trace~$\TensorField{}\equiv\TensorField{_\a^\a}$. The linearisation of general relativity may be extended with a tuned Fierz--Pauli mass term to form the linearisation of massive gravity
\begin{equation}\label{eq:FP}\begin{aligned}
	\mathcal{S}(\Coupling{}) = \int & \mathrm{d}^4x\, \Big[ \Coupling{_1} \Big(
	\tfrac{1}{2} \PD{_\b}\TensorField{} \PD{^\b}\TensorField{}
	- \PD{^\a}\TensorField{_\a_\b} \PD{^\b}\TensorField{} \\
	&\quad - \tfrac{1}{2} \PD{_\g}\TensorField{^\a^\b} \PD{^\g}\TensorField{_\a_\b}
	+ \PD{_\b}\TensorField{^\a^\b} \PD{^\g}\TensorField{_\a_\g}
\Big) \\
&\quad - \Coupling{_2} \Big(
	\TensorField{_\a_\b}\TensorField{^\a^\b} - \TensorField{}^2
\Big) \Big].
\end{aligned}\end{equation}
Note that without canonical normalization of the field, one would typically have
\begin{equation}\label{noncanon_coup}
 \Coupling{_1}=-\MPl^2/4,
 \quad
 \Coupling{_2}=m^2\MPl^2/8,
\end{equation}
where the reduced Planck mass is~$\MPl \approx \SI{2.43e18}{\giga\electronvolt}$ and the square graviton mass is~$m^2=-2\Coupling{_2}/\Coupling{_1}$.

\begin{figure}[htbp]
\centering
\includegraphics[width=0.4\linewidth]{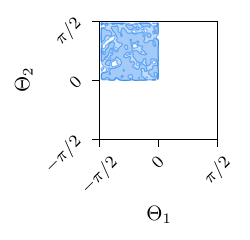}
\caption{The unitary prior~$\pi(\Theta)$ for~\cref{eq:FP}: the uniform measure on the compactified square~$\Theta_i = \tan^{-1}\Coupling{_i}$, restricted to the no-ghost, no-tachyon region and shown by its sample density. Consistent with~\cref{eq:FP_unitarity}. Note that the~$-\pi/2 < \Theta < \pi/2$ range will be assumed in the subsequent~\cref{fig:fpp,fig:tdiff,fig:mv,fig:pheno}.}
\label{fig:fp}
\end{figure}

Since the~$\Coupling{}$ are a priori unconstrained, it is convenient to insist on canonical normalization and to compactify the parameter space to a hypercube of dimension~$N=2$ (i.e., a square), by introducing the notation
\begin{equation}\label{eq:compactification}
	\tensor{\Theta}{_i}\equiv\tan^{-1}\Coupling{_i}/{\mu^{\MassDimension{_i}}},
\end{equation}
for~$\MassDimension{_i}\equiv [\Coupling{_i}]$. Here,~$\mu$ is a reference scale which may be set to unity at this stage (it will become necessary to specify a~$\mu$ in~\cref{results}). From~\cite{Fierz:1939zz,Fierz:1939ix}, it is known that the no-ghost and no-tachyon conditions following from~\cref{eq:FP} are
\be\label{eq:FP_unitarity}
\Coupling{_1} < 0, \quad \Coupling{_2} > 0,
\ee
and a numerical implementation of polology confirms this in~\cref{fig:fp}. This figure is produced by the algorithm whose technical details are deferred to~\cref{implementation} --- here only the general principles are explained.

The unitary region defines a prior~$\pi(\Theta)$ on the compactified coupling space. The compactification in~\cref{eq:compactification} places an implicit standard Cauchy measure on each coupling,~$\mathrm{d}\mu \propto \prod_i \mathrm{d}\Coupling{_i}/(1 + \Coupling{_i}^2)$ --- heavy-tailed, symmetric under~$\Coupling{}\leftrightarrow 1/\Coupling{}$, and uniform in~$\Theta$. The virtue of the hypercube is that this uniform measure is a product, so its projection onto any subset of the~$\Theta$ remains uniform; structure in the marginal panels of~$\pi(\Theta)$ therefore tracks the unitarity boundary itself, rather than any coordinate Jacobian. This follows the standard nested-sampling convention~\cite{Feroz:2007kg}, already central to the early particle physics global fits of Feroz, Hobson and collaborators~\cite{Feroz:2008wr}, and is forfeited by the hyperspherical chart adopted for high-rank theories below (see~\cref{fig:tensor_deep}).

The boundaries of the prior are informed only by the basic requirements of quantum field theory, and thus form an absolute basis for subsequent phenomenological constraints. The availability of~\cref{eq:FP_unitarity} means that~$\pi(\Theta)$ is known analytically, but this will not always be the case when scaling to more complicated models. With this scaling in mind, the distribution~$\pi(\Theta)$ is best encoded numerically by the density of some finite population of samples. These are generated by a sampling algorithm, driven by a loss function which is designed to reward the generation of unitary samples. Depending on the application, the loss function may also select for desirable features in the spectrum, such as particles of specific spin or parity.

In the present case, the samples are uniformly distributed within the unitary region, with contours at~$0.68$ and~$0.95$ confidence levels determined by boundary-corrected kernel density estimation (KDE), as implemented in \anesthetic~\cite{Handley:2019mfs}. These contours are illustrative: the small number of~$\sim 10^4$ samples results in visibly ragged edges, but the square shape of the unitary region is still clear.

\begin{figure}[htbp]
\centering
\includegraphics[width=.6\linewidth]{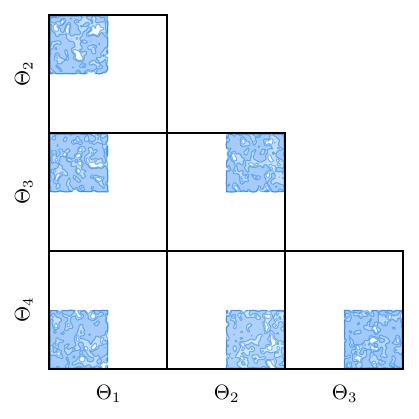}
\caption{Corner plot of the unitary prior~$\pi(\Theta)$ for the Fierz--Pauli--Proca model~\cref{eq:FP,eq:FPP}; the one- and two-dimensional panels are marginal projections of the uniform compactified-hypercube measure restricted to the unitary region. Consistent with~\cref{eq:FPP_unitarity}.}
\label{fig:fpp}
\end{figure}

\paragraph*{Fierz--Pauli--Proca theory} As a simple extension of~\cref{eq:FP}, a massive vector field~$\VectorField{_\a}$ is added via
\begin{equation}\label{eq:FPP}
	\mathcal{S}(\Coupling{}) = \int \mathrm{d}^4x\, \Big[ - \Coupling{_3} \PD{_{[\a}}\VectorField{_{\b]}} \PD{^{[\a}}\VectorField{^{\b]}} - \tfrac{\Coupling{_4}}{2} \VectorField{_\a}\VectorField{^\a} \Big].
\end{equation}
From~\cite{Proca:1936fbw}, the joint unitarity of~\cref{eq:FP,eq:FPP} requires
\be\label{eq:FPP_unitarity}
\Coupling{_1} < 0, \quad \Coupling{_2} > 0, \quad \Coupling{_3} > 0, \quad \Coupling{_4} < 0,
\ee
and these are confirmed in~\cref{fig:fpp}. For the four-parameter model, we see the emergence of the standard `corner-plot' representation of~$\pi(\Theta)$ between all~$\Theta$ pairs; in each case the remaining~$\Theta$ are marginalised.

\begin{figure}[htbp]
\centering
\includegraphics[width=.9\linewidth]{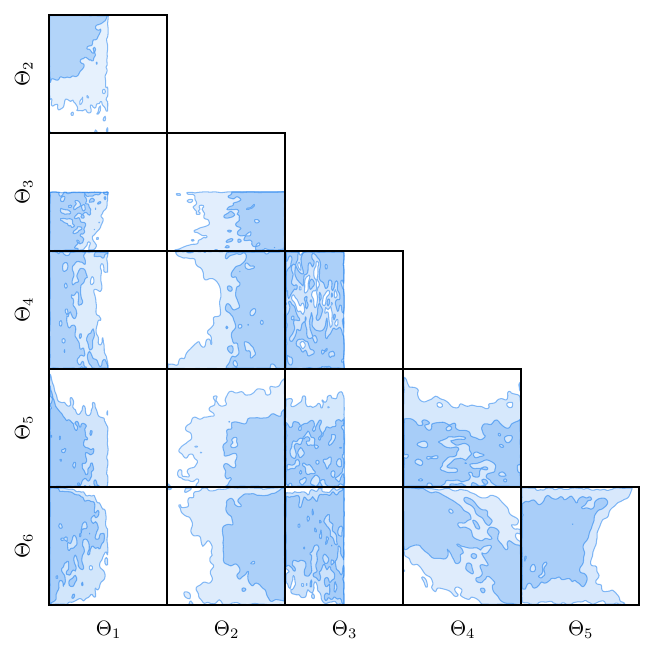}
\caption{Corner plot of the unitary prior~$\pi(\Theta)$ for the Stueckelberg theory~\cref{eq:TDiff}, as in~\cref{fig:fpp}: marginal projections of the uniform compactified-hypercube measure restricted to the no-ghost, no-tachyon region. Consistent with~\cref{eq:TDiff_unitarity}.}
\label{fig:tdiff}
\end{figure}

\paragraph*{Stueckelberg gravity} Whilst~\cref{eq:FP,eq:FPP} are well known models, it is also possible to study new theories. The most general Stueckelberg extension of massive gravity in which transverse diffeomorphism invariance is restored is\footnote{The Fierz--Pauli theory in~\cref{eq:FP} has \emph{no} symmetry, since the mass term breaks the linearised diffeomorphism invariance of the kinetic term. Symmetry may be restored by adding extra Stueckelberg fields~$\VectorField{_\a}$ and~$\ScalarField$, such that~\cref{eq:TDiff} is invariant under the joint transformations~$\TensorField{_\a_\b} \to \TensorField{_\a_\b} + \PD{_{(\a}}\VectorGauge{_{\b)}}$ --- with the transverse constraint~$\PD{_\a}\VectorGauge{^\a} \equiv 0$ --- and~$\VectorField{_\a} \to \VectorField{_\a} + \VectorGauge{_\a} + \PD{_\a}\ScalarGauge$ and~$\ScalarField \to \ScalarField + \ScalarGauge$, where~$\VectorGauge{_\a}$ and~$\ScalarGauge$ are the gauge parameters. Not all the Stueckelberg extensions of massive gravity have been worked out, nor are the implications of the symmetries for consistent interactions fully understood~\cite{Hinterbichler:2011tt,Arkani-Hamed:2002bjr,deRham:2014zqa,Bonifacio:2015rea,deRham:2013qqa,Buchbinder:2012wb}.}
\begin{equation}\label{eq:TDiff}\begin{aligned}
& \mathcal{S}(\Coupling{}) = \int \mathrm{d}^4x\, \Big[ \tfrac{\Coupling{_1}}{2} \TensorField{_\a_\b}\TensorField{^\a^\b}
	+ \Coupling{_5} \TensorField{}^2
	+ \Coupling{_6} \TensorField{} \PD{_\a}\VectorField{^\a} \\
& \quad - \Coupling{_6} \TensorField{} \PD{_\a}\PD{^\a}\ScalarField
	- \Coupling{_1} \PD{_\a}\VectorField{^\a} \PD{_\b}\VectorField{^\b}
	+ \Coupling{_1} \PD{_\b}\VectorField{_\a} \PD{^\b}\VectorField{^\a} \\
& \quad + 2\Coupling{_1} \TensorField{^\a^\b} \PD{_\b}\PD{_\a}\ScalarField
	- 2\Coupling{_1} \TensorField{_\a_\b} \PD{^\b}\VectorField{^\a}
	+ \Coupling{_2} \PD{_\b}\TensorField{} \PD{^\b}\TensorField{} \\
& \quad + \Coupling{_3} \PD{_\a}\TensorField{^\a^\b} \PD{_\g}\TensorField{_\b^\g}
	+ \Coupling{_4} \PD{^\b}\TensorField{} \PD{_\g}\TensorField{_\b^\g} \\
& \quad - \tfrac{\Coupling{_3}}{2} \PD{_\g}\TensorField{_\a_\b} \PD{^\g}\TensorField{^\a^\b} \Big],
\end{aligned}\end{equation}
where~$\VectorField{_\a}$ and~$\ScalarField$ are the Stueckelberg fields.~\cref{fig:tdiff} contains a sizeable unitary volume, and inspection of the pole structure at each point reveals the presence of an additional massive scalar, alongside the massive graviton, whilst there is no propagating spin-one pole. Formally, the bounds of this region can be obtained by a straightforward but lengthy analysis using the techniques of~\cite{Barker:2024juc,Barker:2025qmw}, and are found to be
\be\label{eq:TDiff_unitarity}
\begin{gathered}
\Coupling{_1} < 0, \quad \Coupling{_3} < 0, \quad \Coupling{_5} < \fr{-\Coupling{_1}^2 + \Coupling{_1}\Coupling{_6} - \Coupling{_6}^2}{6\Coupling{_1}}, \\
\Coupling{_2} > \fr{2\Coupling{_1}^2\Coupling{_3} - 2\Coupling{_1}\Coupling{_3}\Coupling{_6} - 6\Coupling{_1}\Coupling{_4}\Coupling{_6} - \Coupling{_3}\Coupling{_6}^2}{12\Coupling{_1}^2}.
\end{gathered}
\ee
Rejection-sampling from~\cref{eq:TDiff_unitarity} can be shown to essentially reproduce the distribution in~\cref{fig:tdiff}, with differences in discrepancy depending on the sampling procedure. The availability of~\cref{eq:TDiff_unitarity} --- whilst reassuring --- calls for expensive algebraic operations. The point is that theories exist only to predict observable phenomena, and a well-explored numerical~$\pi(\Theta)$ should be sufficient for this purpose. We will return to~\cref{fig:tdiff} when astroparticle constraints are applied in~\cref{results}.

\subsection{Untuned theories} \label{untuned}

\paragraph*{Tuning hierarchies} Actually, the theories in~\cref{eq:FP,eq:FPP,eq:TDiff} already have part of the `quantum' prior baked in. This is evident, because in all cases the dimension~$N$ of the~$\Coupling{}$-space is smaller than the number of operators (i.e. the number of terms corresponding to independent index contractions) in~$\mathcal{S}(\Coupling{})$, and this can be viewed as a result of tuning some more general~$\Coupling{}$ to agree with various quantum requirements. A consequence of this tuning is visible in~\cref{fig:fp,fig:fpp,fig:tdiff}, where the unitary regions are seen to occupy a finite \emph{volume} of the remaining~$\Coupling{}$-space.

The~$\Coupling{}$-space of untuned models is not generally expected to contain unitary volumes, but will instead be punctuated by unitary \emph{hypersurfaces} of lower dimension~$d<N$. These hypersurfaces correspond to tuned theories.

\begin{figure}[htbp]
\centering
\includegraphics[width=\linewidth]{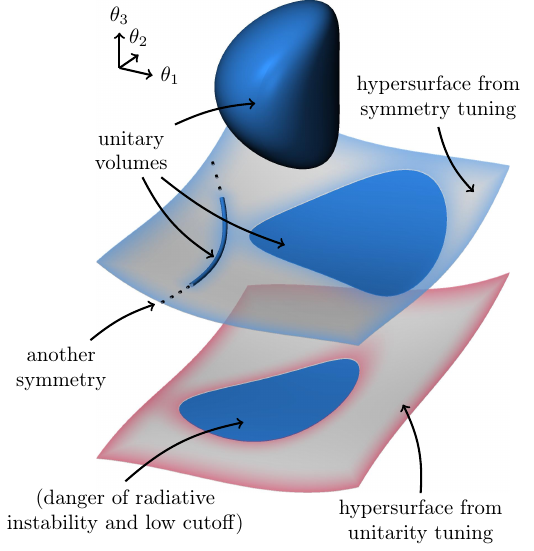}
\caption{More complicated models require the hierarchical identification of unitary theories, which lie on tuned surfaces in~$\Coupling{}$-space. In surfaces defined by unitarity alone, rather than by the emergence of protective gauge symmetries, the validity of the theory may be limited by radiative corrections.}
\label{fig:schematic}
\end{figure}

It is possible to tune for a unitary mass spectrum directly, or to tune for the emergence of new gauge symmetries which happen to admit unitary mass spectra. The two methods are not equivalent, but both can be valid so long as care is taken in the interpretation of the results and -- crucially -- the way in which any~$\mathcal{L}(D|\Coupling{})$ are computed~\cite{Weinberg:1978kz,Coleman:1969sm,Callan:1969sn,Burgess:2007pt}. Symmetry tuning offers stability against radiative corrections, and pushes the cutoff parametrically beyond any native scales when interactions are included. Unitarity tuning typically results in a low cutoff, but the resulting models may still be technically natural EFTs. Tunings can and do occur recursively within sufficiently complex theories. The hierarchy of features that may be identified in this way is illustrated in~\cref{fig:schematic}. It will next be shown that numerical polology is well-suited to finding such features.

\begin{figure*}[!t]
\centering
\includegraphics[width=0.32\textwidth]{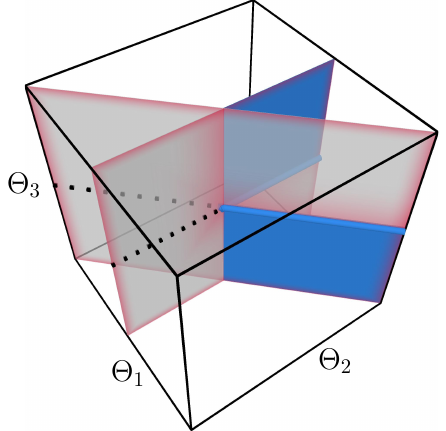}\hfill
\includegraphics[width=0.32\textwidth]{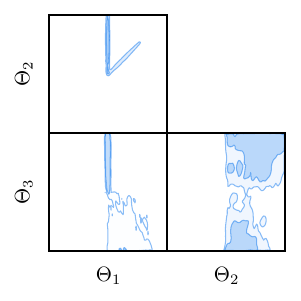}\hfill
\includegraphics[width=0.32\textwidth]{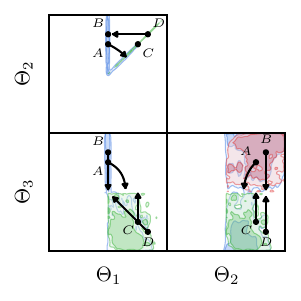}
\caption{The~$N=3$ parameter space of~\cref{eq:MV} and its unitary structure. \emph{Left:} as an example of~\cref{fig:schematic}, a schematic of the compactified hypercube, with the two unitarity-tuned hypersurfaces and the corresponding healthy quadrants highlighted in blue. \emph{Centre:} the unitary~$\pi(\Theta)$ of~\cref{eq:MV}. \emph{Right:} classification of~$\pi(\Theta)$ into Proca (green) and Klein--Gordon-like (red) branches, with the numerical trajectories associated with the emergence of gauge symmetries shown in black. Starting points~$A$ and~$B$ are chosen from the Klein--Gordon-like branch;~$C$ and~$D$ from the Proca branch. Meanwhile, spin-zero symmetries emerge from~$A$ and~$C$, while spin-one symmetries emerge from~$B$ and~$D$. Of these,~$B$ and~$C$ indicate technical naturalness.}
\label{fig:mv}
\end{figure*}

\paragraph*{General vector theory} The simplest illustration involves a vector field~$\VectorField{_\a}$, i.e., 
\begin{equation}\label{eq:MV}\begin{aligned}
    \mathcal{S}(\Coupling{}) = \int \mathrm{d}^4x\, \Big[ &
     \tfrac{\Coupling{_2}}{2} \PD{_\a}\VectorField{^\a} \PD{_\b}\VectorField{^\b}
    -\tfrac{\Coupling{_1}}{2} \PD{_\a}\VectorField{_\b} \PD{^\a}\VectorField{^\b} \\
    & - \tfrac{\Coupling{_3}}{2} \VectorField{_\a}\VectorField{^\a} \Big].
\end{aligned}\end{equation}
The bulk of~$\Coupling{}$-space is sick, but numerical polology cleaves to the healthy regions, as shown in~\cref{fig:mv}. 

In fact,~\cref{fig:mv} is revealing two \emph{separate}~$\mathcal{S}(\Coupling{})$-models, and they can be readily classified. The Proca branch is defined by the tuning~$\Coupling{_1}=\Coupling{_2}$, and is unitary for~$\Coupling{_1} > 0$ and~$\Coupling{_3} < 0$. There exists a second branch defined by the tuning~$\Coupling{_1}=0$, which is unitary for~$\Coupling{_2} > 0$ and~$\Coupling{_3} > 0$. This second theory propagates a massive scalar particle, not a vector.

Both branches are a product of unitarity tuning, and may be classified according to the spins of the massive poles. Neither branch is protected by any gauge symmetry, and so they may also be classified according to technical naturalness. This is done by numerically computing the trajectories in~$\Coupling{}$-space which lead to the emergence of gauge symmetries --- the details are provided in~\cref{implementation}. Points from either branch have two such trajectories, corresponding to the spin sectors (zero or one) in which the symmetries arise. One trajectory tunnels to the opposing branch through the low-$\pi(\Theta)$ void, while the other adheres to the high-$\pi(\Theta)$ surface of the same branch. The latter trajectories correspond to the technically natural deformations of each theory. All trajectories terminate in the massless limit~$\Coupling{_3} =0$. In the Proca branch, the emergent symmetry is that of Maxwell theory; the other branch recovers a (transverse) symmetry of the spin-one sector.\footnote{Specifically, the emergent symmetry is~$\VectorField{_\a} \to \VectorField{_\a} + \VectorGauge{_\a}$, with~$\PD{_\a}\VectorGauge{^\a} \equiv 0$. As shown in~\cite{Barker:2024juc}, this symmetry has the curious effect of removing the massless limit of the spin-zero pole, such that the~$\Coupling{_3}=\Coupling{_1}=0$ theory propagates \emph{nothing}. This differs from the Proca case, where two out of three vector polarizations survive in the Maxwell limit, with the~$U(1)$ symmetry~$\VectorField{_\a} \to \VectorField{_\a} + \PD{_\a}\ScalarGauge$.}

\paragraph*{General tensor theory} The general theory of a single symmetric rank-two tensor field offers a less trivial example than~\cref{eq:MV}, and is given by
\begin{equation}\label{eq:Tensor}\begin{aligned}
	\mathcal{S}(\Coupling{}) = \int \mathrm{d}^4x\, & \bigg[ \Coupling{_1} \TensorField{_\a_\b}\TensorField{^\a^\b}
	+ \Coupling{_2} \TensorField{}^2
	+ \Coupling{_3} \PD{_\b}\TensorField{} \PD{^\b}\TensorField{} \\
& + \Coupling{_4} \PD{_\a}\TensorField{^\a^\b} \PD{_\g}\TensorField{_\b^\g}
	+ \Coupling{_5} \PD{^\b}\TensorField{} \PD{_\g}\TensorField{_\b^\g} \\
& + \Coupling{_6} \PD{_\g}\TensorField{_\a_\b} \PD{^\g}\TensorField{^\a^\b} \bigg].
\end{aligned}\end{equation}
Note that~\cref{eq:Tensor} contains~\cref{eq:FP} as a special case, and like~\cref{eq:TDiff} it has six~$\Coupling{}$. Whilst~\cref{eq:Tensor} has been extensively `mined' for viable tunings in the literature~\cite{VanNieuwenhuizen:1973fi,Rivers:1964nfl,vanderBij:1981ym,Alvarez:2006uu,Bonifacio:2015rea}, an exhaustive analysis is still lacking, and this provides an opportunity to demonstrate the scalability of the method.

\begin{figure*}[!t]
\centering
\includegraphics[width=\textwidth]{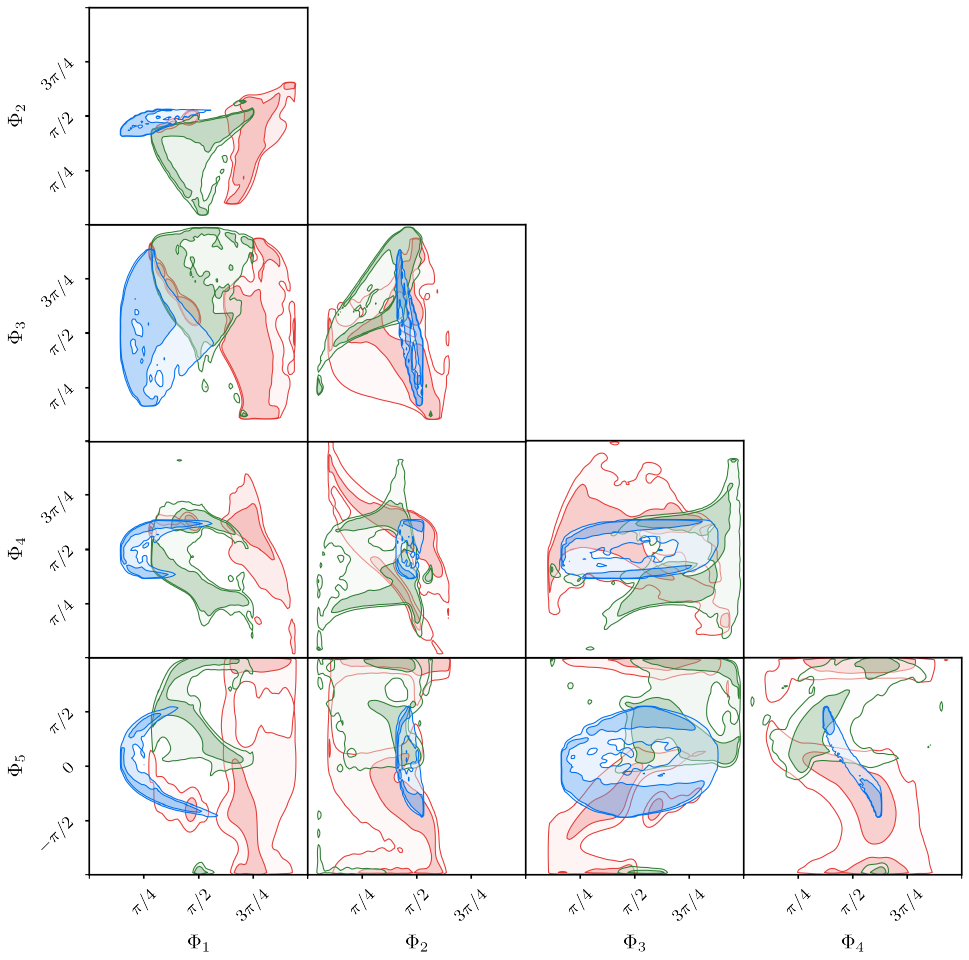}
\caption{`Full-sky' survey of~\cref{eq:Tensor}, which is the most general theory of a symmetric rank-two field~$\TensorField{_\a_\b}\equiv\TensorField{_{(\a\b)}}$, showing branches which propagate one unitary, massive pole, with all other poles hierarchically separated. The Fierz--Pauli family of theories, which propagate a massive spin-two mode, is shown in red; the theory is also capable of propagating a spin-one mode (green) or a spin-zero mode (blue). For this high-resolution survey, the six~$\Coupling{}$ of~\cref{eq:Tensor} are restricted to~$S^5$, covered using the hyperspherical polar coordinates defined in~\cref{rotations}. Unlike the compactified hypercube, the uniform measure on~$S^5$ does not project uniformly, so the displayed density is reweighted to the hyperspherical measure.
}
\label{fig:tensor_deep}
\end{figure*}

For a high-resolution survey, the~$\Theta$ compactification in~\cref{eq:compactification} is wasteful, since the radial direction is redundant under a rescaling of~$\mathcal{S}(\Coupling{})$. It is better to consider the unit-hyperspherical~$S^5$ slice of the~$\Coupling{}$-space, and to cover this `full sky' using the hyperspherical polar coordinates defined in~\cref{rotations}. Since~\cref{eq:Tensor} contains a rich variety of structures, it is practical to limit the survey by adapting the loss function to seek out models which propagate only a single massive mode of any spin: the result is shown in~\cref{fig:tensor_deep}.

Within~\cref{eq:Tensor} there are two such branches which may be written down analytically. The first is defined by the tuning 
\be\label{eq:ScalarTuning}
\Coupling{_4} = \Coupling{_5} = \Coupling{_6} = 0,
\ee
for which the model reduces to Klein--Gordon theory in the trace~$\TensorField{}$, with the trace-free part~$\TensorField{_\a_\b} - \tfrac{1}{4}\TensorField{} \eta_{\a\b}=0$ vanishing on-shell. The conditions in~\cref{eq:ScalarTuning} define a subspace of dimension~$d=3$ in the original~$\Coupling{}$-space, or dimension~$d-1=2$ in the full~$S^5$ `sky', but an application of the techniques in~\cite{Barker:2024juc,Barker:2025qmw} reveals that only the sector 
\be\label{eq:ScalarUnitary}
\Coupling{_1} + 4\Coupling{_2} < 0, \quad \Coupling{_3} > 0,
\ee
is unitary within this surface.

The second evident branch corresponds to Fierz--Pauli theory itself. As defined in~\cref{eq:FP}, it would seem that the two remaining couplings describe a subspace of dimension~$d-1=1$ on the~$S^{5}$ surface, parameterised by some angle~$\varphi$, where
\begin{subequations}\label{eq:FP_phi}
\begin{align}
\sin\varphi &\equiv -\Coupling{_1} = \Coupling{_2}, \\
\cos\varphi &\equiv 2\Coupling{_3} = \Coupling{_4} = -\Coupling{_5} = -2\Coupling{_6}.
\end{align}
\end{subequations}
However, the physical content is invariant under the field redefinition~$\TensorField{_\a_\b} \to \TensorField{_\a_\b} - \sigma\, \TensorField{}\, \eta_{\a\b}$ for most values of~$\sigma$. This trace shift sweeps the surface into a~$d-1=2$ sheet according to the coupling transformations
\begin{subequations}\label{eq:FP_sigma}
\begin{align}
\Coupling{_2} &\to \Coupling{_2} - 2\sigma(\Coupling{_1} + 4\Coupling{_2}) + 4\sigma^2(\Coupling{_1} + 4\Coupling{_2}), \\
\Coupling{_3} &\to \Coupling{_3} - \sigma(8\Coupling{_3} + \Coupling{_5} + 2\Coupling{_6}) \nonumber \\
              &\qquad + \sigma^2(16\Coupling{_3} + \Coupling{_4} + 4\Coupling{_5} + 4\Coupling{_6}), \\
\Coupling{_5} &\to \Coupling{_5} - 2\sigma(\Coupling{_4} + 2\Coupling{_5}),
\end{align}
\end{subequations}
which leave~$\Coupling{_1}$,~$\Coupling{_4}$ and~$\Coupling{_6}$ invariant.

Note that these branches correspond to massive spin-zero and spin-two particles, but numerical polology also reports the existence of a third branch, which propagates a spin-one particle. An analytic investigation with the \emph{Kummitus} software~\cite{Marzo:2026yjg} confirms this discovery: after some experimentation, a unitary spin-one model can be found with the tunings
\begin{equation}\label{eq:vector_branch}
\begin{gathered}
\Coupling{_6} = 0, \quad \Coupling{_1} = 1, \quad \Coupling{_4} = -2, \quad \Coupling{_3} = -\tfrac{1}{8}\Coupling{_5}^2, \\
\Coupling{_2} = \tfrac{1}{12}\big(2\Coupling{_5} - \Coupling{_5}^2 - 4\big),
\end{gathered}
\end{equation}
corresponding to a one-parameter slice of the new branch. A complete analytic understanding of the spin-one branch is still lacking, although it is shown numerically in~\cref{fig:local_dim_histogram} that the branch has effective dimension~$d=3$, in common with the other two branches. Accordingly, the emergence of the spin-one branch marks the point of departure into numerically-driven model discovery.  Other branches of~\cref{eq:Tensor} are expected to exist, with richer multi-particle spectra.

\paragraph*{Rank-three fields} Similar analyses are performed for the general theories of a pair-antisymmetric and a totally symmetric rank-three tensor field
\begin{equation}\label{eq:S123Theory}
	\AntRankThreeField{_\a_\b_\g}\equiv\AntRankThreeField{_\a_{[\b\g]}},\qquad
	\SymRankThreeField{_\a_\b_\g}\equiv\SymRankThreeField{_{(\a\b\g)}}.
\end{equation}
The fields in~\cref{eq:S123Theory} have a natural interpretation as the torsion and non-metricity tensors, which are predicted to exist in a wide class of extensions of general relativity (see e.g.~\cite{Hehl:1994ue}).\footnote{Strictly, the full non-metricity tensor has the smaller symmetry~$\SymRankThreeField{_\a_\b_\g}\equiv\SymRankThreeField{_\a_{(\b\g)}}$, but totally symmetric fields are also important in the study of higher-spin models, see e.g.~\cite{Fronsdal:1978rb,Singh:1974qz}.} Their general actions, analogous to~\cref{eq:Tensor}, are given in~\cref{rank3}. As with~\cref{eq:Tensor}, the torsion-like field admits theories of isolated massive particles with spin zero, one or two; for the non-metricity-like field, there is an additional spin-three branch.

\paragraph*{Effective model-dimension} Under the interpretation of theories as data models, as set out in~\cref{intro}, it is important to understand the effective number of model parameters. In the tuned case, the unitary subvolume simply inherits the dimensionality~$d=N$ of the parent~$\Coupling{}$-space. For untuned theories,~$d$ is a derived parameter.

All the branches derived from all the parent actions in~\cref{untuned} correspond to free theories of a single massive particle, and so it may be expected that the effective~$d=1$ --- the particle mass --- in all cases: this is certainly true when phenomenological constraints are applied to the free theory alone, as will be seen in~\cref{results}. It is shown in~\cref{interactions_main,interactions}, however, that quantum corrections from the interacting theory \emph{also} depend on the free couplings, and that they do so in combinations other than the particle mass formulae.

It thus seems safest to track the full geometric~$d$, and it is discussed in~\cref{implementation-appendix,local_dimensionality} how this can be estimated numerically from the chain of samples that encodes the prior. In~\cref{table:dims}, the effective~$d$ for the discovered branches of the various untuned theories are shown.

\begin{table}[!htbp]
\centering
\caption{Effective integer dimensionality~$d$ for theories of single massive particles of given spin derived from~\cref{eq:MV,eq:Tensor,eq:A23,eq:S123}, read off from~\cref{fig:local_dim_histogram} in~\cref{local_dimensionality}.}
\label{table:dims}
\renewcommand{\arraystretch}{1.3}
\begin{tabular*}{\columnwidth}{@{\extracolsep{\fill}}c|cccc}
\hline\hline
Field & Spin-zero & Spin-one & Spin-two & Spin-three \\
\hline
$\VectorField{_\alpha}$                 & 2 & 2 & --- & --- \\
$\TensorField{_\alpha_\beta}\equiv\TensorField{_{(\alpha\beta)}}$                 & 3 & 3 & 3 & --- \\
$\AntRankThreeField{_\alpha_\beta_\gamma}\equiv\AntRankThreeField{_\alpha_{[\beta\gamma]}}$ & 5 & 4 & 4 & --- \\
$\SymRankThreeField{_\alpha_\beta_\gamma}\equiv\SymRankThreeField{_{(\alpha\beta\gamma)}}$  & 3 & 3 & 3 & 3 \\
\hline\hline
\end{tabular*}
\end{table}

\begin{figure*}[!t]
\centering
\includegraphics[width=0.32\textwidth]{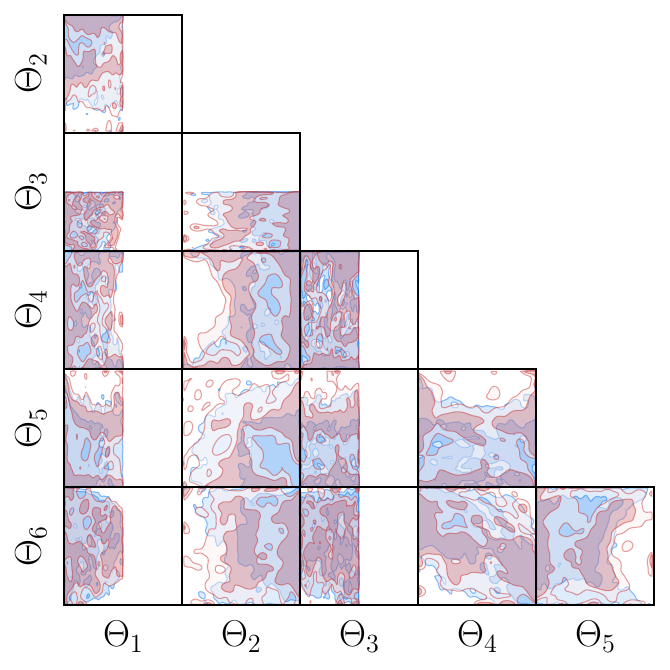}%
\hfill
\includegraphics[width=0.32\textwidth]{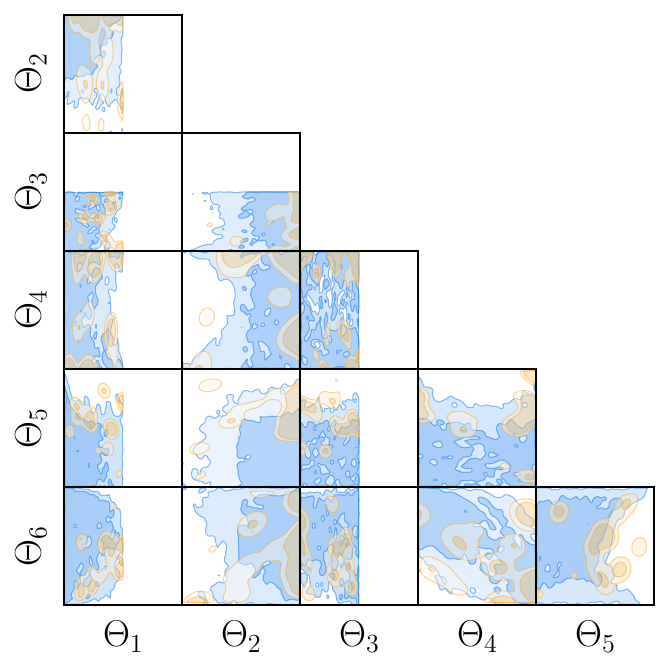}%
\hfill
\includegraphics[width=0.32\textwidth]{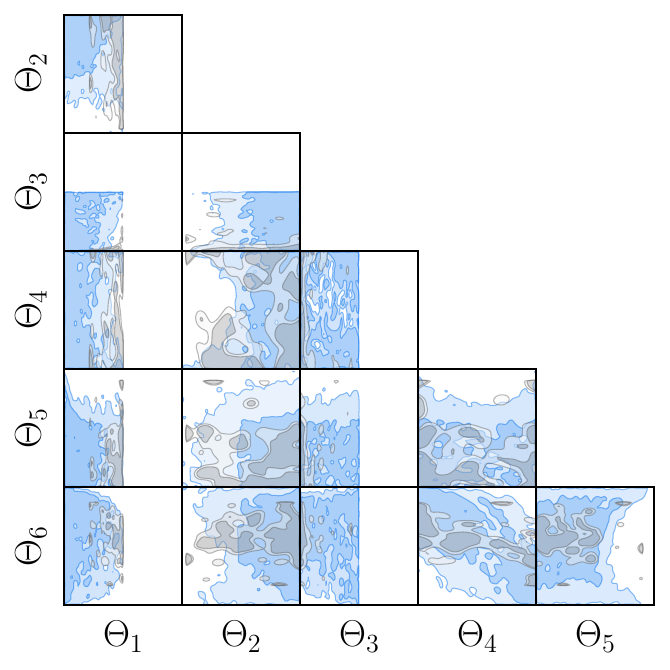}
\caption{Phenomenological constraints on the unitary prior of~\cref{eq:TDiff}. Blue shows the unitarity prior~$\pi(\Theta)$ alone; coloured shows the same prior reweighted by (i.e.\ multiplied by) the indicated observational likelihood. \emph{Left:} black hole superradiance (M33~X-7) applied to the spin-zero particle using the \BHSR survival probability; couplings are~$\tensor{\Theta}{_i}\equiv\tan^{-1}\Coupling{_i}/{\mu^{\MassDimension{_i}}}$, for mass dimension~$\MassDimension{_i}\equiv [\Coupling{_i}]$, where the reference scale is~$\mu = \SI{1.43e-12}{\electronvolt}$. \emph{Middle:} dark energy constraint on the spin-zero particle at scale~$\mu = \SI{1.41e-32}{\electronvolt}$, using the BAO$+$SNe~Ia marginalised likelihood (DESI~DR2$+$Pantheon$+$SH0ES). \emph{Right:} GWTC-3 graviton mass bound~$m\leq \SI{2.42e-23}{\electronvolt}$ on the spin-two particle at scale~$\mu = \SI{1.52e-22}{\electronvolt}$.}
\label{fig:pheno}
\end{figure*}

\subsection{Astroparticle physics} \label{results}

\paragraph*{Tree-level phenomenology} The derived quantities associated with numerical polology admit various phenomenological constraints, which can be applied as multiplicative weights to the `quantum' prior, thereby producing a tighter composite prior.

This is demonstrated using three independent probes, applied to the theory in~\cref{eq:TDiff}. For all probes, derived masses are exploited. This requires us to fix the reference scale~$\mu$ in~\cref{eq:compactification}, i.e., the unit in which the~$\Coupling{}$ of non-zero mass dimension are measured; the computation of mass dimensions is discussed in~\cref{implementation}. Each choice of $\mu$ defines a new prior, which has been tailored to the sensitivity of the probe. Phenomenologically, attention is drawn to mass scales which maximally affect the observables; in astrophysics and cosmology these are very low, at~$m\ll\SI{1}{\electronvolt}$. An appropriate choice of~$\mu$ is close to this scale, and thus shields the numerics from its small value.\footnote{In principle, numerical problems could still arise for untuned theories, for which the priors are obtained by sweeping sick and unwanted poles into a hierarchy --- see~\cref{implementation-appendix}. Increasing this hierarchy is numerically costly, requiring higher-precision arithmetic without which the pathological sector may appear at a phenomenologically relevant scale if~$\mu\ll\SI{1}{\electronvolt}$. To overcome this problem, it suffices to simply neglect the sick poles in phenomenological computations.} Note that the reference scale~$\mu$, as defined in~\cref{eq:compactification}, is a global parameter of the prior; it should not be confused with the mass~$m$ of the particle, which varies locally in the coupling space. In particular, the symbolic techniques of~\cite{Barker:2024juc,Barker:2025qmw} can be applied to~\cref{eq:TDiff}, to show that the spin-zero and spin-two masses are given respectively by
\begin{subequations}\label{eq:TDiff-masses}
\begin{align}
    m^2 &= \frac{-2\,\theta_1\bigl(\theta_1^2 + 6\,\theta_1\theta_5 - \theta_1\theta_6 + \theta_6^2\bigr)}{2\,\theta_1^2(6\,\theta_2 - \theta_3) + 2\,\theta_1(\theta_3 + 3\,\theta_4)\,\theta_6 + \theta_3\,\theta_6^2}, \label{eq:TDiff-mass-spin0}\\
    m^2 &= \frac{\theta_1}{\theta_3}, \label{eq:TDiff-mass-spin2}
\end{align}
\end{subequations}
whilst there is no propagating spin-one pole. Although~\cref{eq:TDiff-masses} are relatively simple expressions, as with~\cref{eq:TDiff_unitarity} this will not always be the case. Indeed, it is pointed out in~\cite{Barker:2025qmw} that closed-form expressions for the mass poles may not exist even for the free theory, due to the Abel--Ruffini theorem. The framework developed in~\cref{implementation} completely bypasses this difficulty, by computing the masses numerically as derived parameters.

The most rigorous cosmological pipelines tend to assume general relativity as the underlying gravitational framework, into which `modules' of new physics are injected. Consequently, the model in~\cref{eq:TDiff} offers an awkward case-study: as a theory of massive gravity, different from general relativity, its eventual non-linear completion may call for arduous pipeline revisions.

Since the following constraints are intended to be illustrative, we will mostly gloss over this limitation (see~\cref{superradiance,darkenergy} for full details). Indeed, due to these and other difficulties, `modified' theories of gravity are not presently the most promising frontier for new physics, whilst the dark matter problem presents as a more useful target.

\paragraph*{Superradiance} Ultralight bosons trigger superradiant instabilities of spinning black holes (BHs), so BH observations exclude boson masses within a corresponding window~\cite{Arvanitaki:2014wva}. In its basic form, superradiance is sensitive only to the Lagrangian itself, and is agnostic with respect to the dynamical configuration of the fields. Constraints arise for any BH whose horizon is comparable to the Compton wavelength of the boson. Heavy species are constrained by X-ray binaries (XRBs) and gravitational wave (GW) events; light species by active galactic nuclei. The BH spin, and thus the quality of the constraint, depends strongly on the population from which the BH is drawn.

Beyond the free mass, superradiant instability is additionally sensitive to the boson spin, and to self-interactions. A thorough superradiance pipeline will thus be able to take advantage of the full tower of derived parameters in numerical polology. This becomes useful if, for example, an improved understanding of BH demography reveals gaps that cannot be explained by astrophysical processes alone. The prior in~\cref{fig:tdiff} is reweighted under the following hypothesis:
\begin{center}
	``\emph{The massive spin-zero particle predicted by~\cref{eq:TDiff} is susceptible to superradiant instabilities near BHs.}''
\end{center}

This is done using observations of M33~X-7~\cite{Orosz:2007ng}, a stellar-mass XRB which naturally probes masses near~$m\sim \SIrange{e-13}{e-11}{\electronvolt}$. The weight is computed using the \BHSR software of Hoof \emph{et al.}~\cite{Hoof:2024quk}, with further details given in~\cref{superradiance}.\footnote{Note that \BHSR does not compute superradiance rates for spins higher than zero, though such an extension does not pose any fundamental technical difficulty; the theory in~\cref{eq:TDiff} does not propagate a spin-one particle, but the spin-two particle could in principle be constrained in this way.} By design, the Bayesian framework set out in~\cite{Hoof:2024quk} is natively extensible to the inclusion of more BH posteriors as such data products become available, and so makes an ideal phenomenological `plug-in' for numerical polology. As shown in~\cref{fig:pheno} (left), superradiance excludes~$63.8\%$ of the prior volume at global reference scale~$\mu = \SI{1.43e-12}{\electronvolt}$. As mentioned above, the reference~$\mu$ is distinct from the actual spin-zero mass~$m$. For this hypothesis, it is effectively the dependence of the mass formula in~\cref{eq:TDiff-mass-spin0} on the underlying couplings, and their compactification via~\cref{eq:compactification} at the scale~$\mu$, that drives the superradiance constraint on the prior. Crucially, however, the closed-form analytic expression for the mass is not needed during the numerical evaluation.

\paragraph*{Dark energy} In the canonical quintessence scenario, a massive scalar field has an effective equation of state that transitions from~$w\approx -1$ to~$w \approx 0$ at the epoch~$m\sim H(z)$~\cite{Turner:1983he,Marsh:2015xka}. Variations on this theme are known to strongly affect the Hubble tension via early dark energy models~\cite{Poulin:2018cxd,Kamionkowski:2022pkx}, and may be useful in response to recent signals for dynamical dark energy at late times~\cite{DESI:2024mwx,DESI:2025zgx}. Such effects can, in principle, be constrained to as far back as big bang nucleosynthesis, beyond which the expansion history becomes speculative. The prior in~\cref{fig:tdiff} is reweighted under the following hypothesis:
\begin{center}
	``\emph{The massive spin-zero particle predicted by~\cref{eq:TDiff} is responsible for 100\% of dark energy.}''
\end{center}

This is done using the DESI~DR2 BAO measurements~\cite{DESI:2025zgx} combined with Pantheon and SH0ES supernovae~\cite{Scolnic:2021amr}. Our pipeline follows the analysis of Ormondroyd \emph{et al.}~\cite{Ormondroyd:2025iaf,Ormondroyd:2025exu}, with further details given in~\cref{darkenergy}. The resulting constraint on the spin-zero mass, shown in~\cref{fig:pheno} (middle), excludes~$90.7\%$ of the prior volume at~$\mu = \SI{1.41e-32}{\electronvolt}$. As with the superradiance example, this reweighting is effectively sensitive only to the formula in~\cref{eq:TDiff-mass-spin0}.

\paragraph*{Graviton dispersion} A massive graviton modifies the dispersion relation of GWs, causing frequency-dependent propagation speeds that are detectable in compact binary coalescence signals. The prior in~\cref{fig:tdiff} is reweighted under the following hypothesis:
\begin{center}
	``\emph{The massive spin-two particle predicted by~\cref{eq:TDiff} is identified as the graviton which mediates GWs.}''
\end{center}

The GWTC-3 combined analysis of 43 binary BH events yields the bound~$m\leq \SI{2.42e-23}{\electronvolt}$ at~$90\%$ credibility~\cite{LIGOScientific:2021sio}. As shown in~\cref{fig:pheno} (right), the GWTC-3 bound excludes~$20.1\%$ of the prior volume at~$\mu = \SI{1.52e-22}{\electronvolt}$. This time, the reweighting is effectively sensitive only to the formula in~\cref{eq:TDiff-mass-spin2}.

In summary, the example of~\cref{eq:TDiff} --- despite being a somewhat arbitrary theory --- demonstrates that the method is well suited to the imposition of observational constraints. We do not extend here to actual model selection, or attempt to combine the three constraints. This is because $\mu$ changes by ten orders of magnitude between the three probes, which motivates the use of three different priors. Note that in rescaling $\mu$, we only update our \emph{interpretation} of the compactified coordinates in~\cref{eq:compactification}, without any Jacobian factor distorting the desired uniform measure.\footnote{The boundaries of the unitary region are defined by the signs of dimensionally consistent functions of the couplings, and so they remain static when~$\mu$ changes.} Consequently, each prior is broadly conditioned so that its structure coincides with the experimental sensitivity. In this scheme, it is unsurprising that reweightings such as those shown in~\cref{fig:pheno} show a high percentage of excluded prior volume; for actual inference applications, this informativeness may be quantified by the Kullback--Leibler divergence. Moreover, there are principled alternatives to the uniform prior measure, such as a maximum-entropy $\mu^2$-scale exponential on $m^2$. For free theories this is well motivated, and straightforward to implement: the likelihoods are only sensitive to the mass spectra, which are already generated as derived parameters in the prior chain. For the interacting case, the problem may be more involved, depending on the experiment. Other principled measures, including those motivated by quantum effects in the presence of interactions (see e.g.~\cite{Fowlie:2023egw}), will be discussed in the companion paper.

\subsection{Interactions}\label{interactions_main}

\paragraph*{Interactions} Many more observational constraints become accessible when interactions are included, so it is essential that the proposed framework accommodates these. As mentioned in~\cref{intro}, the `quantum' prior on the interaction couplings must meet various strict requirements.

Whilst the interacting sector is not the main focus of this work, it is helpful to provide an example through the Fierz--Pauli theory in~\cref{eq:FP}. The field can be canonically normalized with~$\MPl$, so that~\cref{noncanon_coup} is replaced by
\begin{equation}\label{canon_coup}
 \Coupling{_1}=-1/4,
 \qquad
 \Coupling{_2}=m^2/8.
\end{equation}
Consistent interactions for gravity are known to be found in the the Einstein--Hilbert operator: by expanding perturbatively in powers of the field, the cubic and quartic extensions to the part of~\cref{eq:FP} parameterised by~$\Coupling{_1}$ are given in~\cref{eq:TensorDerivativeInteractions} --- these operators have common coefficients of~$\Coupling{_1}/\MPl{}$ and~$\Coupling{_1}/\MPl{}^2$, respectively. For simplicity, it is assumed for the moment that the tuned Fierz--Pauli mass term (the part proportional to~$\Coupling{_2}$) does \emph{not} have any non-linear extensions, i.e. that the massive gravity potential is strictly quadratic.

\paragraph*{Radiative corrections} Working upwards from the free theory in~\cref{eq:FP}, the tree-level two-point function is described by the propagator, and at one loop by the sunset diagram (bilinear in cubic couplings) and the tadpole (linear in quartic couplings):
\begin{center}
\resizebox{\linewidth}{!}{%
\begin{tikzpicture}[baseline={(0,-0.05)}]
\begin{scope}[xshift=0cm]
  \begin{feynman}
    \vertex (a) at (-1.0, 0);
    \vertex (b) at ( 1.0, 0);
    \diagram*{ (a) -- [boson] (b), };
  \end{feynman}
\end{scope}
\begin{scope}[xshift=3.2cm]
  \begin{feynman}
    \vertex (a)  at (-1.5, 0);
    \vertex [dot] (vL) at (-0.6, 0) {};
    \vertex [dot] (vR) at ( 0.6, 0) {};
    \vertex (b)  at ( 1.5, 0);
    \diagram*{
      (a)  -- [boson] (vL),
      (vL) -- [boson, half left, looseness=1.2] (vR),
      (vL) -- [boson, half right, looseness=1.2] (vR),
      (vR) -- [boson] (b),
    };
  \end{feynman}
\end{scope}
\begin{scope}[xshift=6.9cm]
  \begin{feynman}
    \vertex (a)    at (-1.5, 0);
    \vertex [dot] (v) at (0, 0) {};
    \vertex (apex) at (0, 1.2);  
    \vertex (b)    at ( 1.5, 0);
    \diagram*{
      (a) -- [boson] (v) -- [boson] (b),
      (v)    -- [boson, half left,  looseness=1.4] (apex),
      (apex) -- [boson, half left,  looseness=1.4] (v),
    };
  \end{feynman}
\end{scope}
\end{tikzpicture}%
}
\end{center}
The one-loop diagrams lead to shifts in the free couplings of the theory. In general, \emph{all} the free couplings are shifted. With a dimensional regularisation parameter~$\varepsilon$, and using `\dots' to denote finite contributions, we find
\begin{subequations}\label{eq:KineticOnlyShifts}
\begin{align}
\delta\Coupling{_1} &= \frac{55\,\FPCoupling{_2}^2}{36\,\FPCoupling{_1}^2\,\MPl^2\,\pi^2\,\varepsilon} + \dots,  \label{eq:KineticOnlyShift1}\\
\delta\Coupling{_2} &= -\frac{5\,\FPCoupling{_2}^2}{72\,\FPCoupling{_1}^2\,\MPl^2\,\pi^2\,\varepsilon} + \dots, \label{eq:KineticOnlyShift2}\\
\delta\Coupling{_3} &= \frac{35\,\FPCoupling{_2}}{216\,\FPCoupling{_1}\,\MPl^2\,\pi^2\,\varepsilon} + \dots, \label{eq:KineticOnlyShift3}\\
\delta\Coupling{_4} &= -\frac{5\,\FPCoupling{_2}}{36\,\FPCoupling{_1}\,\MPl^2\,\pi^2\,\varepsilon} + \dots, \label{eq:KineticOnlyShift4}\\
\delta\Coupling{_5} &= \frac{5\,\FPCoupling{_2}}{27\,\FPCoupling{_1}\,\MPl^2\,\pi^2\,\varepsilon} + \dots, \label{eq:KineticOnlyShift5}\\
\delta\Coupling{_6} &= \frac{55\,\FPCoupling{_2}}{72\,\FPCoupling{_1}\,\MPl^2\,\pi^2\,\varepsilon} + \dots, \label{eq:KineticOnlyShift6}
\end{align}
\end{subequations}
where the LHS uses the coupling parametrisation of the general tensor theory in~\cref{eq:Tensor}, and on the RHS we use the fresh notation 
\begin{equation}\label{free_redef}
	\Coupling{_1}\to\FPCoupling{_1}, \quad \Coupling{_2}\to\FPCoupling{_2},
\end{equation}
to distinguish the free Fierz--Pauli couplings in~\cref{eq:FP} --- which are used to actually compute the corrections --- from those appearing on the LHS.

\paragraph*{Detuning} \cref{eq:KineticOnlyShifts} illustrates how the introduction of the Fierz--Pauli mass through~$\FPCoupling{_2}$ detunes --- not only the Fierz--Pauli mass term itself --- but also the kinetic structure in the linearization of the Einstein--Hilbert term.

As mentioned in~\cref{untuned}, the tuning of the free theory is critical to the health of the model. In this case, however, the effects vanish along with the graviton mass as~$\FPCoupling{_2}\to 0$, coinciding with the emergence of the gauge symmetry, and this is an indicator of technical naturalness.

Formulae such as those given in~\cref{eq:KineticOnlyShifts} may be readily obtained for more general models. In~\cref{interactions} the corresponding radiative corrections are shown with all possible cubic and quartic algebraic couplings included. Based on such expressions, it is apparent that numerical sampling procedures may be used to reconstruct consistent interactions --- either with strict closure under renormalization, or with technically natural limits --- on top of the tree-level analyses in~\cref{tuned,untuned,results}.

We now proceed in~\cref{implementation} to discuss the technical details of the approach, with an emphasis on the tree-level theory.

\section{Theoretical development}\label{implementation}

\paragraph*{Mass spectra} In this section, we will outline how the priors in~\cref{fig:fp,fig:fpp,fig:tdiff,fig:mv} may be obtained. \cref{fig:tensor_deep,table:dims} require a separate implementation, whose discussion is deferred to~\cref{implementation-appendix}.

As explained in~\cite{Barker:2024juc,Barker:2025qmw}, the technical details of the polology algorithm, as applied to models of the form given in~\cref{EFTLag}, differ between the massive and massless sectors due to the change in the little group. In the massive case, it suffices to work in the representation of spin~$J$ and parity~$P$; in the massless case, a full component decomposition of the fields is required.

Fortunately, our focus on phenomenologically relevant models allows us to avoid the massless sector in our analysis, since additional radiative degrees of freedom are excluded by the thermal history.\footnote{More thoroughly, this is only true for populated species in thermal equilibrium~\cite{Hall:2009bx,Bernal:2017kxu}.} At every numerically sampled~$\Coupling{}$, we thus work with the full suite of massive poles, and seek to navigate the~$\Coupling{}$-space on the basis of their physical health. Beyond this restriction, we will also limit our analysis to parity-preserving theories.\footnote{The algorithm for parity-violating theories has been worked out in~\cite{Barker:2025qmw}, based on work by Karananas in~\cite{Karananas:2014pxa}, and may be applied in future work.}

\paragraph*{Symmetries} Focussing on the sector of spin~$J$ and parity~$P$, the full collection of terms~$\sum_i \Coupling{_i}\,\Operator{i}$ corresponding to the free sector of~\cref{EFTLag} give rise to a Hermitian wave operator, which can be written in a basis of spin-projection operators as~$\WaveOperatorJP{J}{P}$, where~$k\equiv\sqrt{\tensor{k}{_\a} \tensor{k}{^\a}}$ is the momentum. For general~$k$, this admits a collection of null eigenvectors~$\WaveOperatorJP{J}{P}\cdot\NullVector{J}{P}{a} = 0$ which encode all the gauge symmetries acting on the~$J^P$ sector.

Once the~$\theta$ are sampled, these symmetries may be computed numerically from a polynomial expansion
\bea\label{eq:PolyExpansions}
\WaveOperatorJP{J}{P} &= \sum_{n} \WaveOperatorJPn{J}{P}{n} k^n, \\
\NullVector{J}{P}{a} &= \sum_{n} \NullVectorN{J}{P}{a}{n} k^n,
\eea
which gives rise to a block singular value decomposition problem, as follows. Truncating the ansatz~$\NullVectorN{J}{P}{a}{n}$ at degree~$N_{J^P}\,(\dim\WaveOperatorJP{J}{P}-1)$, where~$N_{J^P}$ is the highest non-vanishing order in~$\WaveOperatorJP{J}{P}$, yields a finite linear system~$\BlockMatrix{J}{P} \cdot \BlockNullVector{J}{P}{a} = 0$, where~$\BlockNullVector{J}{P}{a}$ stacks the~$\NullVectorN{J}{P}{a}{n}$, and where the block matrix~$\BlockMatrix{J}{P}$ is built from the~$\WaveOperatorJPn{J}{P}{n}$ in a Toeplitz pattern
\be\label{eq:BlockMatrix}
\BlockMatrix{J}{P} \equiv \begin{pmatrix}
\WaveOperatorJPn{J}{P}{0} & \cdots & 0 \\
\vdots & & \vdots \\
\WaveOperatorJPn{J}{P}{N_{J^P}} & \ddots & 0 \\
0 & & \WaveOperatorJPn{J}{P}{0} \\
\vdots & & \vdots \\
0 & \cdots & \WaveOperatorJPn{J}{P}{N_{J^P}}
\end{pmatrix}.
\ee
The genuine~$\NullVector{J}{P}{a}$ may then be reassembled from the solution for~$\BlockNullVector{J}{P}{a}$.

\paragraph*{Masses and tachyons} Once the symmetries have been found, the regularised wave operator
\be\label{eq:RegularizedWaveOperator}
\RegWaveOperatorJP{J}{P} \equiv \WaveOperatorJP{J}{P} + \sum_a \NullVector{J}{P}{a} \cdot \NullVectorConj{J}{P}{a},
\ee
may be inverted, and this inverse is proportional to the propagator. The poles of the propagator correspond to the physical states, and must also correspond to the roots of~$\det\RegWaveOperatorJP{J}{P}$.

Numerically, a candidate spectrum of poles can be found without symbolic manipulation via a polynomial eigenvalue problem in~$k$ --- though~\cref{implementation-appendix} presents an alternative approach based on the Vandermonde and Frobenius methods. Because the construction in~\cref{eq:PolyExpansions} does not give rise to normalised~$\NullVector{J}{P}{a}$, this spectrum will generally be polluted by spurious `gauge' poles, though these are readily stripped by comparing their null eigenvectors with the gauge sector of the kernel. The physical poles are associated with null eigenvectors, denoted by
\be\label{eq:PhysNullVector}
\RegWaveOperatorJPms{J}{P}{s} \cdot \PhysNullVector{J}{P}{s} = 0.
\ee
The poles occur in pairs, because~$\det\RegWaveOperatorJP{J}{P}$ is formally known to be a polynomial in~$k^2$; for each zero of this polynomial, the principal second root is identified with the mass~$\Mass{J}{P}{s}$ of the corresponding pair of poles. This naturally extends the convention of positive masses for real poles to the complex case, which now have positive real part. In general, only real poles are healthy, since
\be\label{eq:NoTachyon}
\Mass{J}{P}{s}^2 > 0,
\ee
is the no-tachyon criterion. 

\paragraph*{Residues and ghosts} The no-ghost criterion depends on the pole residue
\be\label{eq:MassiveNoGhost}
\Residue{J}{P}{s} > 0,
\ee
which was shown in~\cite{Barker:2024juc} to be given by
\be\label{eq:ResidueDefinition}
\Residue{J}{P}{s} \equiv \NewRes{k^2}{\Mass{J}{P}{s}^2}\left(P\,\tr\,\MoorePenroseJP{J}{P}\right),
\ee
where~$\MoorePenroseJP{J}{P}$ is the Moore--Penrose pseudoinverse of~$\WaveOperatorJP{J}{P}$. Even with sampled~$\Coupling{}$, obtaining the pseudoinverse remains a symbolic problem. A numerical alternative follows from the matrix~$\GaugeMatrixJP{J}{P}$ whose columns are the~$\NullVectorm{J}{P}{a}{s}$, and which defines the non-gauge projection
\bea\label{eq:GaugeProjection}
\ProjNullVector{J}{P}{s} &\equiv \left[ \mathsf{1} - \GaugeProjector{J}{P} \right] \cdot \PhysNullVector{J}{P}{s}, \\
\GaugeProjector{J}{P} &\equiv \GaugeMatrixJP{J}{P} \cdot \left( \GaugeMatrixJPConj{J}{P} \cdot \GaugeMatrixJP{J}{P} \right)^{-1} \cdot \GaugeMatrixJPConj{J}{P},
\eea
of the massive null eigenvectors. For real, positive~$\Mass{J}{P}{s}$, the residue in~\cref{eq:ResidueDefinition} is then given by the expansion
\bea\label{eq:NoGhost}
\Residue{J}{P}{s} = 2P \Bigg[ & \sum_{n}  n \, \left|\Mass{J}{P}{s}\right|^{n-2} \\
				      & \times \ProjNullVectorConj{J}{P}{s} \cdot \WaveOperatorJPn{J}{P}{n} \cdot \ProjNullVector{J}{P}{s}\Bigg]^{-1},
\eea
where the modulus is introduced to provide a simple analytic continuation to complex poles, though the interpretation as the residue is lost in this case. Since~$\WaveOperatorJPn{J}{P}{n}$ is Hermitian for all~$n$, so~$\Residue{J}{P}{s}$ is always real by construction.

\paragraph*{Mass dimensions} So far, the mass dimensions of the~$\Coupling{}$ have not been specified. Indeed, the fields themselves only acquire a definite canonical dimension once the spectrum has been computed, and the same must be true of the~$\Coupling{}$. Let~$\MassDimension{_i}\equiv [\Coupling{_i}]$, whilst~$[\Mass{J}{P}{s}]\equiv 1$. By this point in the calculation, the~$\Mass{J}{P}{s}$ are numerically available, and a rescaling of the mass unit~$\mu\to\beta\mu$ so that~$\Coupling{_i}\to\beta^{\MassDimension{_i}}\Coupling{_i}$, and~$\Mass{J}{P}{s}\to\beta\Mass{J}{P}{s}$, yields
\be\label{eq:MassRescaling}
\sum_{i} \MassDimension{_i} \, \frac{\partial\ln \Mass{J}{P}{s}}{\partial \ln \Coupling{_i}} = 1.
\ee
Any choice of~$\MassDimension{_i}$ that satisfies~\cref{eq:MassRescaling} for all~$\Mass{J}{P}{s}$ is valid. To ensure that the Lagrangian density has the correct mass dimension, under the usual convention that propagating bosons have a canonical dimension of unity, integer solutions are sought with~$\min_i \MassDimension{_i} = 0$.

It is this control over the mass dimensions that allows us to impose the phenomenological reference scale~$\mu$ used in~\cref{results}.

\paragraph*{Technical naturalness} Returning to the question of gauge symmetries, the singular values~$\BlockSV{J}{P}{a}$ of~$\BlockMatrix{J}{P}$ as defined in~\cref{eq:BlockMatrix} provide a useful diagnostic for the proximity of the theory to a gauge-symmetric configuration. To navigate towards such a surface from a given sample~$\Coupling{}$, we perform steepest descent on a chosen~$\BlockSV{J}{P}{a}$. The gradient is provided by the singular-value analogue of the Hellmann--Feynman theorem
\be\label{eq:SVHellmannFeynman}
\frac{\partial \BlockSV{J}{P}{a}}{\partial \Coupling{_i}} = \Re\left[\BlockLSVConj{J}{P}{a} \cdot \frac{\partial \BlockMatrix{J}{P}}{\partial \Coupling{_i}} \cdot \BlockNullVector{J}{P}{a}\right],
\ee
where~$\BlockLSV{J}{P}{a}$ and~$\BlockNullVector{J}{P}{a}$ are the corresponding left and right singular vectors of~$\BlockMatrix{J}{P}$. Because each~$\WaveOperatorJPn{J}{P}{n}$ is linear in the couplings, the derivatives may be precomputed.

\paragraph*{Sampling likelihood} So far, it was only described how to numerically obtain physical quantities from a given sample~$\Coupling{}$. It is now necessary to efficiently generate samples of the prior distribution~$\pi(\Coupling{})$. For performant sampling of high-$N$~$\Coupling{}$-space, it is appropriate to use a nested sampler~\cite{Skilling:2004pqw,Skilling:2006gxv} (see implementations in~\cite{Feroz:2007kg,Handley:2015fda,cabezas2024blackjax,yallup2026nested}). An additional advantage of nested sampling is its natural integration with the Bayesian framework set out in~\cref{eq:evidence}. In this context, the loss function mentioned in~\cref{tuned} itself corresponds to a likelihood.

There are many ways to construct a heuristic likelihood for quantum field theory.\footnote{The posterior of one experiment is the prior of the next. In this framework, model-building begins with the empirically established laws of quantum field theory, which give rise to~$\Likelihood$. Subsequent experiments give rise to the~$\mathcal{L}(D|\Coupling{})$ of~\cref{eq:evidence}. Note that a \emph{uniform} prior is assumed when working with~$\Likelihood$, and the usual ambiguity remains in finding appropriate coordinates for~$\Coupling{}$-space.} Focussing on the tuned theories of~\cref{tuned}, for which the spectra are already known, the goal is to find the sub-volume of the~$\Coupling{}$-space which is unitary. Since unitarity is boolean --- either present or not --- it is necessary to define a `fuzzy' continuation by means of which the sampler is driven by a unitarity-violation penalty; this can be done with the choice 
\begin{multline}\label{eq:Likelihood}
\log\Likelihood = \sum_{J,P} \sum_{s_{J^P}}
 \Bigl[ \mathcal{U}\!\left(\Residue{J}{P}{s}\right)
	\\
        + \mathcal{U}\!\left(-\left|\Im\Mass{J}{P}{s}\right|\right) \Bigr].
\end{multline}
\cref{eq:Likelihood} is a sum over contributions from all the poles, in which the function~$\mathcal{U}$ acts on~$\Residue{J}{P}{s}$ and~$m(\Coupling{})$, since they share the same mass dimension. According to~\cref{eq:NoTachyon,eq:MassiveNoGhost}, the central requirement of~$\mathcal{U}$ is sensitivity to the sign of its argument. Work done by the nested sampler in the healthy region is always wasted, and so we adopt~$\mathcal{U}(x)=0$ for~$x> 0$. By the same logic, such plateaus should be \emph{avoided} in the sick region.

The transition between sick and healthy regions is not always via the origin:~$\Residue{J}{P}{s}$ and~$\Mass{J}{P}{s}$ vary according to a (potentially very complicated) problem in algebraic geometry, which means that changes of sign may also occur across singularities in~$\Coupling{}$-space. Consequently, infinite unitarity violation must be mapped smoothly back to zero; this can be achieved with
\be\label{eq:Compression}
\mathcal{U}(x) \equiv \begin{cases}
	0, & x > 0, \\
	-\dfrac{2 \, |x|^u}{1 + |x|^{2u}}, & x \leq 0,
\end{cases}
\ee
where~$u$ is a global parameter which can assist convergence in cases where the power-law near a sign change is extreme --- we adopt~$u=0.5$ in this work.

The construction in~\cref{eq:Likelihood} ensures that the likelihood is maximised at~$\Likelihood=1$ by unitarity. Nested sampling does not maximise the likelihood, but instead reflects the distribution of its volume. It is thus necessary to filter out the~$\Likelihood<1$ samples in the final chain. This is the procedure that leads to the prior distributions of tuned theories shown in~\cref{fig:fp,fig:fpp,fig:tdiff}, and it is also effective for the simple untuned theory shown in~\cref{fig:mv}.

For more complex untuned theories, the case is altered (see~\cref{implementation-appendix} for an example implementation): the final samples must land on set-of-measure-zero hypersurfaces, and so root-finding becomes central to a successful algorithm. It is imagined that the application to untuned theories will be far more useful than in the tuned case, since this allows hitherto unknown theories to be discovered.

\section{Conclusions} \label{discussion}

\paragraph*{Overview} The literature contains a large and growing plethora of new physics models. By mutual exclusion, most (perhaps all) of these theories cannot be realised in nature, and the theoretical return on investment seems especially low given the historical rate at which the standard models of particle physics and cosmology have actually been updated. More precisely, the problem is not that there are \emph{too many} models, but rather that the techniques for constructing them have not been adequately mechanised.

This work proposes a framework for systematically exploring theory-space, using \emph{numerical polology}. The method is restricted to perturbative bosonic degrees of freedom. Within these bounds, the need for theoretical innovation is minimised, since quantum field theory already provides the complete set of principles (no more, no less) needed for the development of predictive theories. Sampling is central to the method, allowing it to integrate naturally with the techniques of model selection in precision cosmology. In particular, the unitary~$\pi(\Coupling{})$ is a principled, field-theoretic prior over the free couplings. For simple theories, Bayesian global fits already place explicit priors on such couplings (see e.g.~\cite{GAMBIT:2021rlp,Balazs:2022tjl,Chang:2022jgo,Chang:2023cki}), typically as generic (flat or logarithmic) ranges supplemented by theory-validity cuts (from perturbativity, EFT-validity and partial-wave unitarity). Numerical polology systematically extends this approach to high-dimensional coupling spaces, including those for which the self-consistent theories are not yet known. The approach focuses on the quantum prior~$\pi(\Coupling{})$, leaving the likelihood~$\mathcal{L}(D|\Coupling{})$ to the established machinery.

\paragraph*{Outlook} The approach has several limiting factors. Principally, we do not consider the effects of curvature, or background values of the fields. Whilst this hinders some cosmological applications, it does not preclude them, since background values can often be substituted in phenomenological likelihood pipelines. Superradiance provides a good example of this, since the formulae in~\cref{superradiance} were derived in Kerr spacetime.

On the other hand, it can be dangerous to transplant effective field theories between regimes in this way. Programmes already exist for systematically constructing models in the early~\cite{Cheung:2007st,Weinberg:2008hq} and late~\cite{Creminelli:2008wc,Gubitosi:2012hu,Gleyzes:2013ooa} Universe, where the Hubble flow alters the symmetry and introduces a scale around which the effective theory is organised. The validity of the dark energy likelihood in~\cref{darkenergy} may be limited by these considerations.

As a possible future approach to cosmological backgrounds, note that the spin-projection operator formalism of~\cite{VanNieuwenhuizen:1973fi,Neville:1978bk,Neville:1979rb,Sezgin:1981xs,Sezgin:1979zf,Kuhfuss:1986rb,Karananas:2014pxa,Karananas:2016ltn,Mendonca:2019gco,Percacci:2020ddy,Marzo:2021esg,Marzo:2021iok,Mikura:2023ruz,Mikura:2024mji} --- upon which the present method relies --- was recently extended to maximally symmetric spacetimes~\cite{Hutchings:2024qqf}. Similarly, the technical details of the algorithm have been worked out only for bosonic fields, so (e.g.) fermionic dark matter is not yet accessible: the extension to fermions is likely to be straightforward, because spin-projection operators admit half-integer representations~\cite{Isaev:2017nud}.

Lastly, whilst the possibility of numerically computing quantum corrections from interactions has been demonstrated (see~\cref{interactions_main,interactions}), the assembly of the full interaction likelihood is left to the companion paper. Apart from extending the dimensionality of the coupling space, the inclusion of interactions and the corresponding renormalization group may motivate a more principled \emph{measure} on the quantum prior~\cite{Fowlie:2023egw}. This extends the compactified hypercube or hyperspherical measures used in the present work, which has focused primarily on the problem of validity bounds at tree level.

\paragraph*{Data availability} The sources and chains used in this work are made available in the supplemental materials at~\cite{supp_materials}.

\begin{acknowledgments}
We are grateful for useful discussions with Justin Feng, Mark Hindmarsh, Sebastian Hoof, Yoann Launay, Roberto Percacci, Fethi Ramazano\v{g}lu, Syksy R\"as\"anen, Ignacy Sawicki, Richard Woodard and David Yallup.

This work was supported by the research environment and infrastructure of the Handley Lab at the University of Cambridge.

W.~B. is grateful for the hospitality of the Helsinki Institute of Physics and the Cavendish Laboratory at the University of Cambridge, and was supported by Marie Sklodowska-Curie Actions

C.~M. acknowledges support by the Estonian Research Council grant PRG1677 and the CoE program TK202 `\emph{Fundamental Universe}'. 

This work used the DiRAC Data Intensive service~(CSD3 \href{www.csd3.cam.ac.uk}{www.csd3.cam.ac.uk}) at the University of Cambridge, managed by the University of Cambridge University Information Services on behalf of the STFC DiRAC HPC Facility~(\href{www.dirac.ac.uk}{www.dirac.ac.uk}). The DiRAC component of CSD3 at Cambridge was funded by BEIS, UKRI and STFC capital funding via STFC capital grants ST/P002307/1 and ST/R002452/1 and STFC operations grant ST/R00689X/1. DiRAC is part of the UKRI Digital Research Infrastructure.

\paragraph*{Disclaimer} Co-funded by the European Union (Physics for Future – Grant Agreement No. 101081515). Views and opinions expressed are however those of the author(s) only and do not necessarily reflect those of the European Union or European Research Executive Agency. Neither the European Union nor the granting authority can be held responsible for them.
\end{acknowledgments}

\appendix

\section{Rotation matrix} \label{rotations}

\paragraph*{Rotation matrix} Whilst a natural choice for covering~$S^5$, the hyperspherical polar chart is punctuated by coordinate singularities which lie along the axes of the original coordinates. Since these singularities coincide with the physical hypersurfaces in~\cref{eq:ScalarTuning,eq:FP_phi,eq:FP_sigma}, it is desirable to avoid them by rotating the original coordinates. The (random) rotation matrix is given by
\begin{equation}\label{eq:Q}
\mathsf{Q} \equiv \setlength{\arraycolsep}{2pt}\renewcommand{\arraystretch}{0.9}\small\begin{pmatrix}
+0.13 & -0.62 & +0.64 & +0.19 & -0.33 & -0.22 \\
+0.05 & -0.18 & -0.04 & -0.93 & -0.31 & +0.08 \\
+0.03 & +0.71 & +0.68 & -0.12 & -0.12 & +0.06 \\
+0.37 & +0.07 & +0.03 & -0.22 & +0.46 & -0.78 \\
-0.18 & -0.27 & +0.36 & -0.19 & +0.76 & +0.40 \\
+0.90 & -0.01 & -0.05 & +0.08 & +0.03 & +0.42
\end{pmatrix},
\end{equation}
which is used to define~$\tensor{\vartheta}{_i} \equiv \sum_j\tensor{\mathsf{Q}}{_{ij}} \Coupling{_j}/\mu^{\MassDimension{_j}}$, and the hyperspherical polar coordinates are then given as usual by
\begin{subequations}\label{eq:yFromPhi}
\begin{align}
\vartheta_1 &\equiv \cos\Phi_1, \\
\vartheta_2 &\equiv \sin\Phi_1\,\cos\Phi_2, \\
\vartheta_3 &\equiv \sin\Phi_1\,\sin\Phi_2\,\cos\Phi_3, \\
\vartheta_4 &\equiv \sin\Phi_1\,\sin\Phi_2\,\sin\Phi_3\,\cos\Phi_4, \\
\vartheta_5 &\equiv \sin\Phi_1\,\sin\Phi_2\,\sin\Phi_3\,\sin\Phi_4\,\cos\Phi_5, \\
\vartheta_6 &\equiv \sin\Phi_1\,\sin\Phi_2\,\sin\Phi_3\,\sin\Phi_4\,\sin\Phi_5.
\end{align}
\end{subequations}
Further details are available at~\cite{supp_materials}.

\section{Rank-three fields} \label{rank3}

\paragraph*{Rank-three fields} The general parity-preserving actions for free theories built from the fields defined in~\cref{eq:S123Theory} are
\begin{subequations}\label{eq:RankThreeActions}
\begin{align}
	\mathcal{S}(\Coupling{}) = \int \mathrm{d}^4x\, & \bigg[
	  \Coupling{_1} \AntRankThreeField{_\a_\b_\g} \AntRankThreeField{^\a^\b^\g}
	+ \Coupling{_2} \AntRankThreeField{^\a^\b^\g} \AntRankThreeField{_\b_\a_\g} \nonumber \\
	&- \Coupling{_3} \AntRankThreeField{^\b} \AntRankThreeField{_\b}
	 - \Coupling{_4} \PD{_\b} \AntRankThreeField{_\g} \PD{^\g} \AntRankThreeField{^\b} \nonumber \\
	&- \Coupling{_5} \PD{_\g} \AntRankThreeField{_\b} \PD{^\g} \AntRankThreeField{^\b}
	 + \Coupling{_6} \PD{_\b} \AntRankThreeField{^\a^\b^\g} \PD{^\d} \AntRankThreeField{_\a_\g_\d} \nonumber \\
	&+ \Coupling{_7} \PD{_\a} \AntRankThreeField{^\a^\b^\g} \PD{^\d} \AntRankThreeField{_\b_\g_\d}
	 + \Coupling{_8} \PD{_\b} \AntRankThreeField{^\a^\b^\g} \PD{^\d} \AntRankThreeField{_\g_\a_\d} \nonumber \\
	&+ \Coupling{_9} \PD{^\g} \AntRankThreeField{^\b} \PD{^\d} \AntRankThreeField{_\g_\b_\d}
	 + \Coupling{_{10}} \PD{_\a} \AntRankThreeField{^\a^\b^\g} \PD{^\d} \AntRankThreeField{_\d_\b_\g} \nonumber \\
	&+ \Coupling{_{11}} \PD{^\d} \AntRankThreeField{_\a_\b_\g} \PD{_\d} \AntRankThreeField{^\a^\b^\g} \nonumber \\
	&+ \Coupling{_{12}} \PD{^\d} \AntRankThreeField{_\b_\a_\g} \PD{_\d} \AntRankThreeField{^\a^\b^\g}
	\bigg], \label{eq:A23} \\
	\mathcal{S}(\Coupling{}) = \int \mathrm{d}^4x\, & \bigg[
	  \Coupling{_1} \SymRankThreeField{_\a_\b_\g} \SymRankThreeField{^\a^\b^\g}
	+ \Coupling{_2} \SymRankThreeField{^\b} \SymRankThreeField{_\b} \nonumber \\
	&+ \Coupling{_3} \PD{_\b} \SymRankThreeField{_\g} \PD{^\g} \SymRankThreeField{^\b}
	 + \Coupling{_4} \PD{_\g} \SymRankThreeField{_\b} \PD{^\g} \SymRankThreeField{^\b} \nonumber \\
	&+ \Coupling{_5} \PD{_\a} \SymRankThreeField{^\a^\b^\g} \PD{^\d} \SymRankThreeField{_\b_\g_\d}
	 + \Coupling{_6} \PD{^\g} \SymRankThreeField{^\b} \PD{^\d} \SymRankThreeField{_\b_\g_\d} \nonumber \\
	&+ \Coupling{_7} \PD{^\d} \SymRankThreeField{_\a_\b_\g} \PD{_\d} \SymRankThreeField{^\a^\b^\g}
	\bigg], \label{eq:S123}
\end{align}
\end{subequations}
where~$\AntRankThreeField{_\b} \equiv \AntRankThreeField{^\a_\a_\b}$ and~$\SymRankThreeField{_\b} \equiv \SymRankThreeField{^\a_\a_\b}$. If either of these fields are propagating in our Universe, it is reasonable to expect that their free limits are captured by~\cref{eq:RankThreeActions}.

\section{Model dimensionality} \label{local_dimensionality}

\paragraph*{Participation ratio} The model dimensionality on the~$S^{N-1}$ hypersphere, i.e.~$d-1$, is approximated for each posterior sample~$i$ as the participation ratio of the principal singular values of the neighbouring samples.

\begin{figure*}[!htbp]
\centering
\includegraphics[width=\textwidth]{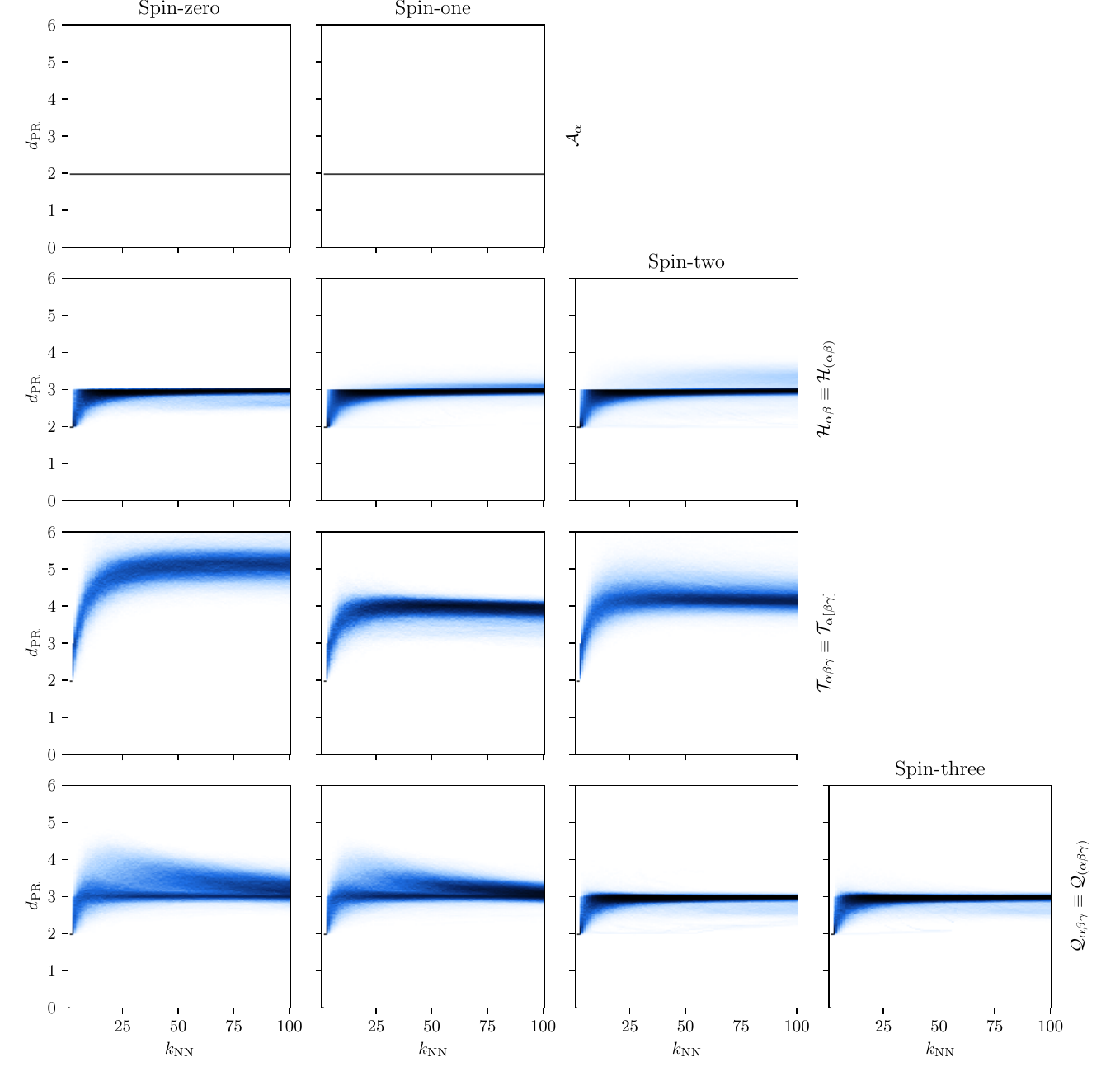}
\caption{Effective model dimensionality of the branches in~\cref{eq:MV,eq:Tensor,eq:A23,eq:S123}. The estimates stabilise with increasing number of nearest neighbours, with noise arising from the curvature of the hypersurfaces in the embedding hyperspheres. The asymptotic values are read off in~\cref{table:dims}.}
\label{fig:local_dim_histogram}
\end{figure*}

Concretely, for a chosen neighbourhood size~$\kNN$ we form the~$\kNN\times N$ matrix~$\mathsf{X}(i, \kNN)$ whose rows are the~$\kNN$ nearest neighbours of sample~$i$ (in the~$N$-dimensional~$\Coupling{}$-space of the parent action, for which the~$\Coupling{}$ are Cartesian coordinates), and take its singular-value decomposition~$\mathsf{X}(i, \kNN) - \overline{\mathsf{X}}(i, \kNN) \equiv \mathsf{U}(i, \kNN)\cdot \mathrm{diag}\!\big(s_a(i, \kNN)\big)\cdot\mathsf{V}(i, \kNN)^{\text{T}}$. The local intrinsic dimension is then defined by
\begin{equation} \label{eq:participation_ratio}
d_{\text{PR}}(i, \kNN) \;\equiv\; \frac{\big(\sum_a s_a(i, \kNN)^2\big)^{\!2}}{\sum_a s_a(i, \kNN)^4}+1.
\end{equation}
The~$+1$ correction on the RHS of~\cref{eq:participation_ratio} ensures that the result corresponds to the dimensionality of the hypersurface in the full~$N$-dimensional~$\Coupling{}$-space, rather than the slice seen on the~$S^{N-1}$ surface that contains the chain. This process is performed for a range of~$\kNN$ values, with the resulting histograms shown in~\cref{fig:local_dim_histogram}.

\paragraph*{Density estimation} Because unitarity is a boolean property, it is important that the prior distribution vanishes in the sick region, whilst being uniform within the healthy region. Uniformity, however, must be defined with respect to some underlying set of coordinates.

For untuned theories, the algorithm in~\cref{implementation-appendix} samples on the~$S^{N-1}$ `sky', and so it is natural to enforce uniformity with respect to the hyperspherical measure. The reweighting is performed by inverting a~$k$NN density estimate. For a chain that occupies a sub-manifold of intrinsic dimension~$d$ inside the~$N$-dimensional~$\Coupling{}$-space, the local density at sample~$i$ is proportional to~$\kNN / r(i, \kNN)^d$, where~$r(i, \kNN)$ is the distance from sample~$i$ to its~$\kNN$ nearest neighbour. The importance weight is then the inverse density~$r(i, \kNN)^d$.

\begin{figure}[!t]
\centering
\includegraphics[width=\columnwidth]{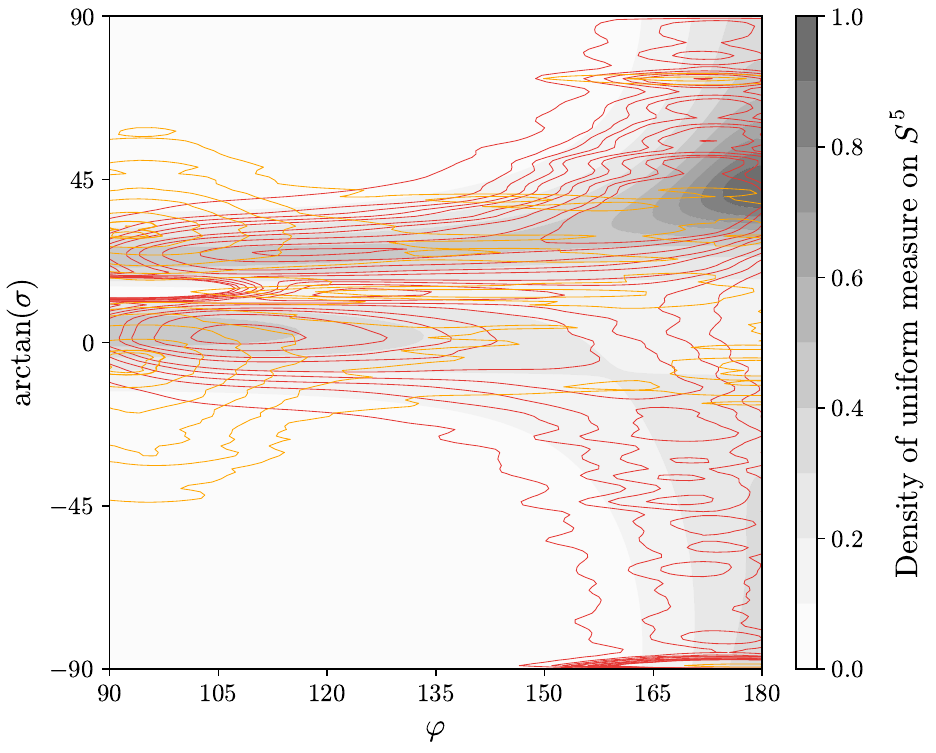}
\caption{Uniform `sky-coverage' of the spin-two (Fierz--Pauli) branch of~\cref{eq:Tensor}, restricted to the analytic unitary portion of the two-dimensional sheet parameterised by the angle~$\varphi$ of~\cref{eq:FP_phi} and the trace-shift~$\sigma$ of~\cref{eq:FP_sigma}. The coloured contours are obtained as KDEs (orange for the unweighted chain, red for the final, reweighted chain); grey contours indicate the raw density of the hyperspherical measure.}
\label{fig:tensor_branches}
\end{figure}

The single integer exponent~$d$ may be approximated by~$d_{\text{PR}}(i, \kNN)$, as given in~\cref{eq:participation_ratio} (specifically, with the choice~$\kNN=20$). \cref{fig:tensor_branches} illustrates, for the spin-two (i.e. Fierz--Pauli) branch of~\cref{eq:Tensor}, that the reweighted chain adheres to the hyperspherical measure as intended. Further details are available at~\cite{supp_materials}.

\section{Superradiance pipeline} \label{superradiance}

\paragraph*{Theoretical development} Following the framework of Hoof \emph{et al.}~\cite{Hoof:2024quk},\footnote{There are important notational differences between~\cite{Hoof:2024quk} and the present work; the former uses~$\mu$ to denote the boson mass, and~$m$ to denote the magnetic quantum number -- the latter uses~$m$ and~$\mL$ for these quantites respectively, whilst~$\mu$ is the global reference scale defined in~\cref{eq:compactification}.} the weight applied to each posterior sample of spin-zero mass~$m$ is the `survival probability' that the chosen BH's measured mass and spin is compatible with the existence of an ultralight scalar of that mass. The stellar-mass X-ray binary M33~X-7~\cite{Liu:2008tk} is modelled as a Kerr BH; the data comprise 1838 joint posterior samples of the BH mass~$\MBH$ and (dimensionless) angular momentum~$J$, expressed in terms of the spin parameter~$\aStar\equiv 8\pi\MPl^2\,J/\MBH^2$.\footnote{As a limitation, we inherit these samples from the \BHSR package of~\cite{Hoof:2024quk}, which obtained them by digitising a figure in~\cite{Liu:2008tk}.}

In the absence of interactions, the scalar mode near the BH must have Klein--Gordon dynamics, and should be well described by the collection of quantum states of the form~$\left\lvert n,l,\mL \right\rangle$, where~$n$,~$l$ and~$\mL$ are the principal, orbital and magnetic quantum numbers, respectively. Superradiant instability corresponds to an exponential growth in the state occupation number~$N\approx \exp\left(\GammaSR t\right)$, where~$t$ is the time elapsed and~$\GammaSR$ is the superradiance rate.

For fixed~$l$, we assume~$\GammaSR$ to be maximised when
\begin{equation} \label{eq:sr_nlm}
n=l+1,\qquad \mL=l.
\end{equation}
Each of~$1\leq l \leq 6$ is compared with the dimensionless gravitational coupling
\begin{equation} \label{eq:sr_alpha}
\alpha \;\equiv\; \frac{\MBH\,m}{8\pi\,\MPl^2},
\end{equation}
in which~$\MPl \approx \SI{2.43e18}{\giga\electronvolt}$ is the reduced Planck mass. If~$\alpha/l > 1/2$ the bound-state assumptions underlying the analytic approximations to~$\GammaSR$ break down, and we conservatively assume no growth (i.e.\ skip the state).

Otherwise, we use the approximations of~\cite{Bao:2022hew}, i.e. 
\begin{equation} \label{eq:sr_gammaSR}
\begin{aligned}
\GammaSR \;\approx\; & y\,z\,c_{l}\,\omega_1\!\prod_{k=1}^{l}\big[k^2 (1 - \aStar^2) + y^2\big], \\
y \;\equiv\; & l\,\aStar - 2\,r_+\,\omega_0, \\
z \;\equiv\; & \big[\alpha^2\,(1 - \omega_0^2/m^2)\big]^{l + 1/2},
\end{aligned}
\end{equation}
where~$r_+ \equiv \big(\MBH/(8\pi\,\MPl^2)\big)(1 + \sqrt{1 - \aStar^2})$ is the outer-horizon radius of the Kerr spacetime in Boyer--Lindquist coordinates. The leading and subleading bound-state frequencies~$\omega_0$ and~$\omega_1$ in~\cref{eq:sr_gammaSR} are
\begin{subequations} \label{eq:sr_omegas}
\begin{align}
\omega_0 \;\equiv \; & m\,\bigg[1 - 2\alpha^2 \Big[(l+1)^2 + 4\alpha^2 \nonumber \\
& \qquad + (l+1)\sqrt{(l+1)^2 + 8\alpha^2}\Big]^{-1}\bigg]^{\!1/2}, \\
\omega_1 \;\equiv \; & \frac{m^2 - \omega_0^2}{(l+1)\,\omega_0}\,\bigg[1 + \frac{4\alpha^2(2\omega_0^2 - m^2)}{m^2(l+1)^2}\bigg]^{-1},
\end{align}
\end{subequations}
while the combinatorial prefactor is
\begin{equation} \label{eq:sr_cnl}
c_{l} \;\equiv\; \frac{2^{4l + 2}\,(l!)^2}{(2l + 1)!\,(2l)!^{\,2}}.
\end{equation}
Note that the maximum-rate conditions of~\cref{eq:sr_nlm} have been substituted into~\cref{eq:sr_gammaSR,eq:sr_omegas,eq:sr_cnl}, whilst the formulae available in~\cite{Bao:2022hew} are more general.

\paragraph*{Reweighting} A given state~$l$ is taken to appreciably modify the BH characteristics over the BH timescale~$\tBH$ if and only if
\begin{equation} \label{eq:sr_growth}
\GammaSR\,\tBH \;>\; \log \NFin,
\end{equation}
where the e-folding budget~$\log\NFin$ corresponds, somewhat arbitrarily, to the extraction of a dimensionless spin~$\Delta\aStar = 0.1$ away from the BH and into the growing boson `cloud'. This budget was found in~\cite{Arvanitaki:2014wva} to be
\begin{equation} \label{eq:sr_nfin}
\NFin \;\equiv\; 10^{76}\,\Big(\frac{\Delta\aStar}{0.1}\Big)\,\Big(\frac{\MBH}{10\,M_\odot}\Big)^{\!2}\,\frac{1}{l}.
\end{equation}
The estimated age of the M33~X-7 system is taken as~$\tBH = \SI{3}{\mega\year}$ (see e.g.~\cite{Gou:2011nq,Hoof:2024quk}).

A posterior sample is judged incompatible with the boson if \emph{any} state satisfies the growth criterion~\cref{eq:sr_growth}. The weight applied to each sample is then the relative fraction of the surviving posterior samples, which can be interpreted as the survival probability of the BH. Further details are available at~\cite{supp_materials}.

\section{Dark energy pipeline} \label{darkenergy}

\paragraph*{Theoretical development} A homogeneous scalar field of mass~$m$ in a cosmological background obeys a damped Klein--Gordon equation\footnote{Note that the scalar~$\phi$ appearing in~\cref{eq:KG_FRW} represents the final propagating degree of freedom in the particle spectrum: it cannot be immediately identified with the field appearing in~\cref{eq:TDiff}, since the propagating mode may contain spin-zero contributions from the higher-rank fields in the action.}
\begin{equation} \label{eq:KG_FRW}
\ddot{\phi} + 3H\dot{\phi} + V'(\phi) = 0,
\qquad V(\phi) \equiv \frac{1}{2} m^2 \phi^2,
\end{equation}
where~$H$ is the Hubble number, so that the discriminant~$9H^2 - 4m^2$ partitions the dynamics into an overdamped (frozen) regime for~$m < 3H/2$ and an underdamped (oscillating) regime for~$m > 3H/2$~\cite{Turner:1983he,Marsh:2015xka}. The scalar field has density and pressure given by 
\begin{equation} \label{eq:rho_p_scalar}
\RhoDe \equiv \frac{1}{2} \dot{\phi}^2 + V(\phi),
\qquad
\PDe   \equiv \frac{1}{2} \dot{\phi}^2 - V(\phi).
\end{equation}
In the overdamped regime we assume slow-roll, i.e.
\begin{equation} \label{eq:slow_roll_phidot}
\dot{\phi} \;\approx\; -\,\frac{V'(\phi)}{3H} \;=\; -\,\frac{m^2 \phi}{3H},
\end{equation}
and substituting~\cref{eq:slow_roll_phidot} into~\cref{eq:rho_p_scalar} yields the corresponding equation of state
\begin{equation} \label{eq:w_slowroll}
\WDe \;\equiv \; \frac{\PDe}{\RhoDe} \;=\; \frac{m^2 - 9H^2}{m^2 + 9H^2},
\qquad (m < 3H/2).
\end{equation}
Meanwhile, for~$m > 3H/2$ the field undergoes coherent oscillations and a WKB time-average gives the matter-like result~$\WDe = 0$~\cite{Turner:1983he}. For this analysis, the two regimes are stitched
\begin{equation} \label{eq:w_eff}
\WDe(m, H) \equiv
\begin{cases}
\dfrac{m^2 - 9H^2}{m^2 + 9H^2}, & m < 3H/2, \\[6pt]
0, & m \ge 3H/2,
\end{cases}
\end{equation}
though a more thorough time-averaged integration would avoid the jump at the transition.

The dark energy density obeys the continuity equation~$\RhoDeDot + 3H(1 + \WDe)\RhoDe = 0$, with formal solution
\begin{equation} \label{eq:f_de}
\begin{aligned}
\FDe(z, m) &\;\equiv\; \frac{\RhoDe(z; m)}{\RhoDe(0; m)} \\
&\;=\; \exp\!\left[\,3\!\int_0^z\!\mathrm{d}z'\,\frac{1 + \WDe(m, H(z'))}{1 + z'}\,\right],
\end{aligned}
\end{equation}
reducing to~$\FDe \to 1$ in the~$\Lambda$CDM limit. Strictly,~$H(z')$ in the integrand depends on~$\FDe$ itself; we evaluate it on a fiducial~$\Lambda$CDM background~$\OmegaMFid \equiv 0.315$ and~$\HZeroFid \equiv \SI{67.4}{\km\per\second\per\mega\parsec}$ (see~\cite{Planck:2018vyg}), such that
\begin{equation} \label{eq:H_fid}
H_{\text{fid}}(z) \equiv \HZeroFid \sqrt{\OmegaMFid(1 + z)^3 + (1 - \OmegaMFid)}.
\end{equation}
Plugging~$\FDe$ from~\cref{eq:f_de} into the first Friedmann equation at the sampled~$\OmegaM$ gives the model's Hubble rate
\begin{equation} \label{eq:E_z}
\begin{aligned}
E&(z; \OmegaM, m) \;\equiv\; \frac{H(z)}{H_0} \\
&\;=\; \sqrt{\OmegaM(1 + z)^3 + (1 - \OmegaM)\,\FDe(z, m)}.
\end{aligned}
\end{equation}

\paragraph*{Reweighting} Once the input in~\cref{eq:E_z} is computed, the remainder of the pipeline is adapted from Ormondroyd \emph{et al.}~\cite{Ormondroyd:2025iaf,Ormondroyd:2025exu}. The three BAO distance ratios reported by DESI in~\cite{DESI:2025zgx} are
\begin{subequations} \label{eq:bao_distances}
\begin{align}
\frac{D_H(z)}{\SoundHorizonDrag} &\equiv \frac{1}{H_0\SoundHorizonDrag\,E(z)}, \\
\frac{D_M(z)}{\SoundHorizonDrag} &\equiv \frac{1}{H_0\SoundHorizonDrag}\!\!\int_0^z\!\!\frac{\mathrm{d}z'}{E(z')}, \\
\frac{D_V(z)}{\SoundHorizonDrag} &\equiv \left(z\,\Big(\frac{D_M}{\SoundHorizonDrag}\Big)^{\!2}\frac{D_H}{\SoundHorizonDrag}\right)^{\!1/3},
\end{align}
\end{subequations}
in which~$\SoundHorizonDrag$ is the comoving sound horizon at the baryon-drag epoch, and~$H_0\SoundHorizonDrag$ appears as a multiplicative nuisance parameter. For Pantheon and SH0ES~\cite{Scolnic:2021amr}, the observable is the apparent magnitude
\begin{equation} \label{eq:dl}
\begin{aligned}
\mB(z_{\text{hd}}, z_{\text{hel}}) \;\equiv\; & 5\,\log_{10}\!\!\left[\,(1 + z_{\text{hel}})\!\int_0^{z_{\text{hd}}}\!\frac{\mathrm{d}z'}{E(z')}\,\right] \\
& + \mathcal{M},
\end{aligned}
\end{equation}
where~$z_{\text{hd}}$ is the redshift corrected for the peculiar motions of the source and observer,~$z_{\text{hel}}$ is the heliocentric redshift, and~$\mathcal{M}$ is a constant calibration offset.

The DESI data gives thirteen values for the quantities on the LHS of~\cref{eq:bao_distances}, spanning seven values of~$z$, together with a covariance matrix~$\CovBAO$. For any values of~$m$,~$\OmegaM$ and~$H_0\SoundHorizonDrag$ we build the model residual vector~$\mathsf{r}$ from the RHS of~\cref{eq:bao_distances} and compute the log-likelihood
\begin{equation} \label{eq:logL_bao}
\log \LikBAO(m, \OmegaM, H_0\SoundHorizonDrag) \equiv -\frac{1}{2}\ \mathsf{r}^{\text{T}}\cdot \CovBAO^{-1}\cdot \mathsf{r}.
\end{equation}

For the supernovae, we take the cut of 1371 entries on the LHS of~\cref{eq:dl} for which~$z_{\text{hd}}>0.023$, so as to exclude observations that might be swept up in the local flow. The predictions are made using only the first term on the RHS of~\cref{eq:dl}, since the offset~$\mathcal{M}$ is unknown and must be marginalised analytically. The log-likelihood is then
\begin{equation} \label{eq:logL_sne}
	\log \LikSNe(m, \OmegaM) \equiv -\frac{1}{2}\ \mathsf{r}^{\text{T}}\cdot \CovSNeProj^{-1}\cdot \mathsf{r},
\end{equation}
where the \emph{projected} inverse covariance is defined by
\begin{equation} \label{eq:invcov_tilde}
\CovSNeProj^{-1} \;\equiv\; \CovSNe^{-1} \;-\; \frac{\CovSNe^{-1}\cdot\OneVec\cdot\OneVec^{\text{T}}\cdot\CovSNe^{-1}}{\OneVec^{\text{T}}\cdot\CovSNe^{-1}\cdot\OneVec},
\end{equation}
and~$\OneVec$ is the all-ones column vector --- the construction in~\cref{eq:invcov_tilde} is `desensitized' to the global calibration shift, and can be viewed as the sought analytic marginalisation of~$\mathcal{M}$.

The remaining nuisance parameters in~\cref{eq:logL_bao,eq:logL_sne} are~$\OmegaM$ and~$H_0\SoundHorizonDrag$. These are integrated out on a generous (see again~\cite{Planck:2018vyg})~$50 \times 50$ grid covering~$\OmegaM \in [0.01, 0.99]$ and~$H_0\SoundHorizonDrag \in [\SI{3650}{\km\per\second}, \SI{18250}{\km\per\second}]$, by 
\begin{align}
\log \LikMar(m) &\;\equiv\; \log\!\!\sum_{i,j} w_{ij}\,\exp\Big[\log \LikSNe\big(m, \OmegaM^{(i)}\big) \nonumber \\
&+ \log \LikBAO\big(m, \OmegaM^{(i)}, (H_0\SoundHorizonDrag)^{(j)}\big)\Big], \label{eq:logL_marg}
\end{align}
where~$w_{ij}$ contains trapezoidal weights and grid spacings.

The dark energy weight is~$\exp\!\big[\Delta\log \Lik(m)\big]$, where
\begin{equation} \label{eq:delta_logL}
\Delta\log \Lik(m) \;\equiv\; \log \LikMar(m) - \log \LikMarLCDM,
\end{equation}
and the~$\Lambda$CDM reference is the same marginalisation as in~\cref{eq:logL_marg}, but evaluated with~$\WDe = -1$ identically, i.e.~$\FDe = 1$. Further details are available at~\cite{supp_materials}.

\section{Interactions} \label{interactions}

\paragraph*{Derivative interactions} By perturbatively expanding the Einstein--Hilbert term, the cubic and quartic interactions which build on top of the free kinetic sector of~\cref{eq:FP} are found to be
\begin{subequations}\label{eq:TensorDerivativeInteractions}
\begin{align}
    &\mathcal{S}(\Coupling{}) = \frac{\Coupling{_1}}{\MPl} \int \mathrm{d}^4x\, \bigg[ -\tfrac{3}{2} \TensorField{^\a^\b} \PD{_\a}\TensorField{^\g^\d} \PD{_\b}\TensorField{_\g_\d} \nonumber\\
    &\quad +\tfrac{1}{2} \TensorField{^\a^\b} \PD{_\a}\TensorField{} \PD{_\b}\TensorField{} -2 \TensorField{^\a^\b} \PD{_\b}\TensorField{} \PD{_\g}\TensorField{_\a^\g} \nonumber\\
    &\quad -2 \TensorField{^\a^\b} \PD{_\b}\TensorField{_\a^\g} \PD{_\g}\TensorField{} -2 \TensorField{_\a^\g} \TensorField{^\a^\b} \PD{_\g}\PD{_\b}\TensorField{} \nonumber\\
    &\quad + \TensorField{} \TensorField{^\b^\g} \PD{_\g}\PD{_\b}\TensorField{} +2 \TensorField{_\a^\g} \TensorField{^\a^\b} \PD{_\g}\PD{_\d}\TensorField{_\b^\d} \nonumber\\
    &\quad - \TensorField{} \TensorField{^\b^\g} \PD{_\g}\PD{_\d}\TensorField{_\b^\d} + \TensorField{^\a^\b} \PD{_\g}\TensorField{} \PD{^\g}\TensorField{_\a_\b} \nonumber\\
    &\quad -\tfrac{1}{4} \TensorField{} \PD{_\g}\TensorField{} \PD{^\g}\TensorField{} +2 \TensorField{^\a^\b} \PD{_\g}\TensorField{_\a^\g} \PD{_\d}\TensorField{_\b^\d} \nonumber\\
    &\quad +4 \TensorField{^\a^\b} \PD{_\b}\TensorField{_\a^\g} \PD{_\d}\TensorField{_\g^\d} - \TensorField{} \PD{_\b}\TensorField{^\b^\g} \PD{_\d}\TensorField{_\g^\d} \nonumber\\
    &\quad -2 \TensorField{^\a^\b} \PD{^\g}\TensorField{_\a_\b} \PD{_\d}\TensorField{_\g^\d} + \TensorField{} \PD{^\g}\TensorField{} \PD{_\d}\TensorField{_\g^\d} \nonumber\\
    &\quad +2 \TensorField{^\a^\b} \TensorField{^\g^\d} \PD{_\d}\PD{_\b}\TensorField{_\a_\g} -2 \TensorField{^\a^\b} \TensorField{^\g^\d} \PD{_\d}\PD{_\g}\TensorField{_\a_\b} \nonumber\\
    &\quad +2 \TensorField{_\a^\g} \TensorField{^\a^\b} \PD{_\d}\PD{_\g}\TensorField{_\b^\d} - \TensorField{} \TensorField{^\b^\g} \PD{_\d}\PD{_\g}\TensorField{_\b^\d} \nonumber\\
    &\quad -\tfrac{1}{2} \TensorField{_\a_\b} \TensorField{^\a^\b} \PD{_\d}\PD{_\g}\TensorField{^\g^\d} +\tfrac{1}{4} \TensorField{}^2 \PD{_\d}\PD{_\g}\TensorField{^\g^\d} \nonumber\\
    &\quad -2 \TensorField{_\a^\g} \TensorField{^\a^\b} \PD{_\d}\PD{^\d}\TensorField{_\b_\g} + \TensorField{} \TensorField{^\b^\g} \PD{_\d}\PD{^\d}\TensorField{_\b_\g} \nonumber\\
    &\quad +\tfrac{1}{2} \TensorField{_\a_\b} \TensorField{^\a^\b} \PD{_\d}\PD{^\d}\TensorField{} -\tfrac{1}{4} \TensorField{}^2 \PD{_\d}\PD{^\d}\TensorField{} \nonumber\\
    &\quad +2 \TensorField{^\a^\b} \PD{_\b}\TensorField{_\g_\d} \PD{^\d}\TensorField{_\a^\g} + \TensorField{^\a^\b} \PD{_\g}\TensorField{_\b_\d} \PD{^\d}\TensorField{_\a^\g} \nonumber\\
    &\quad -3 \TensorField{^\a^\b} \PD{_\d}\TensorField{_\b_\g} \PD{^\d}\TensorField{_\a^\g} -\tfrac{1}{2} \TensorField{} \PD{_\g}\TensorField{_\b_\d} \PD{^\d}\TensorField{^\b^\g} \nonumber\\
    &\quad +\tfrac{3}{4} \TensorField{} \PD{_\d}\TensorField{_\b_\g} \PD{^\d}\TensorField{^\b^\g} \bigg], \label{eq:TensorDerivativeCubic} \\
    &\mathcal{S}(\Coupling{}) = \frac{\Coupling{_1}}{\MPl^2} \int \mathrm{d}^4x\, \bigg[ - \TensorField{^\a^\b} \TensorField{^\g^\d} \PD{_\b}\TensorField{_\d_\e} \PD{_\g}\TensorField{_\a^\e} \nonumber\\
    &\quad +\tfrac{3}{2} \TensorField{_\a^\g} \TensorField{^\a^\b} \PD{_\b}\TensorField{^\d^\e} \PD{_\g}\TensorField{_\d_\e} -\tfrac{3}{4} \TensorField{} \TensorField{^\b^\g} \PD{_\b}\TensorField{^\d^\e} \PD{_\g}\TensorField{_\d_\e} \nonumber\\
    &\quad -\tfrac{1}{2} \TensorField{_\a^\g} \TensorField{^\a^\b} \PD{_\b}\TensorField{} \PD{_\g}\TensorField{} +\tfrac{1}{4} \TensorField{} \TensorField{^\b^\g} \PD{_\b}\TensorField{} \PD{_\g}\TensorField{} \nonumber\\
    &\quad +2 \TensorField{_\a^\g} \TensorField{^\a^\b} \TensorField{^\d^\e} \PD{_\g}\PD{_\b}\TensorField{_\d_\e} -2 \TensorField{_\a^\g} \TensorField{^\a^\b} \TensorField{^\d^\e} \PD{_\g}\PD{_\e}\TensorField{_\b_\d} \nonumber\\
    &\quad +2 \TensorField{_\a^\g} \TensorField{^\a^\b} \PD{_\g}\TensorField{} \PD{_\d}\TensorField{_\b^\d} - \TensorField{} \TensorField{^\b^\g} \PD{_\g}\TensorField{} \PD{_\d}\TensorField{_\b^\d} \nonumber\\
    &\quad +3 \TensorField{^\a^\b} \TensorField{^\g^\d} \PD{_\g}\TensorField{_\a^\e} \PD{_\d}\TensorField{_\b_\e} -2 \TensorField{^\a^\b} \TensorField{^\g^\d} \PD{_\b}\TensorField{_\a^\e} \PD{_\d}\TensorField{_\g_\e} \nonumber\\
    &\quad +2 \TensorField{^\a^\b} \TensorField{^\g^\d} \PD{_\b}\TensorField{_\a_\g} \PD{_\d}\TensorField{} - \TensorField{^\a^\b} \TensorField{^\g^\d} \PD{_\g}\TensorField{_\a_\b} \PD{_\d}\TensorField{} \nonumber\\
    &\quad +2 \TensorField{_\a^\g} \TensorField{^\a^\b} \PD{_\g}\TensorField{_\b^\d} \PD{_\d}\TensorField{} - \TensorField{} \TensorField{^\b^\g} \PD{_\g}\TensorField{_\b^\d} \PD{_\d}\TensorField{} \nonumber\\
    &\quad +2 \TensorField{_\a^\g} \TensorField{^\a^\b} \TensorField{_\b^\d} \PD{_\d}\PD{_\g}\TensorField{} - \TensorField{} \TensorField{_\b^\d} \TensorField{^\b^\g} \PD{_\d}\PD{_\g}\TensorField{} \nonumber\\
    &\quad -\tfrac{1}{2} \TensorField{_\a_\b} \TensorField{^\a^\b} \TensorField{^\g^\d} \PD{_\d}\PD{_\g}\TensorField{} +\tfrac{1}{4} \TensorField{}^2 \TensorField{^\g^\d} \PD{_\d}\PD{_\g}\TensorField{} \nonumber\\
    &\quad -2 \TensorField{_\a^\g} \TensorField{^\a^\b} \TensorField{_\b^\d} \PD{_\d}\PD{_\e}\TensorField{_\g^\e} + \TensorField{} \TensorField{_\b^\d} \TensorField{^\b^\g} \PD{_\d}\PD{_\e}\TensorField{_\g^\e} \nonumber\\
    &\quad +\tfrac{1}{2} \TensorField{_\a_\b} \TensorField{^\a^\b} \TensorField{^\g^\d} \PD{_\d}\PD{_\e}\TensorField{_\g^\e} -\tfrac{1}{4} \TensorField{}^2 \TensorField{^\g^\d} \PD{_\d}\PD{_\e}\TensorField{_\g^\e} \nonumber\\
    &\quad - \TensorField{_\a^\g} \TensorField{^\a^\b} \PD{_\d}\TensorField{} \PD{^\d}\TensorField{_\b_\g} +\tfrac{1}{2} \TensorField{} \TensorField{^\b^\g} \PD{_\d}\TensorField{} \PD{^\d}\TensorField{_\b_\g} \nonumber\\
    &\quad +\tfrac{1}{8} \TensorField{_\a_\b} \TensorField{^\a^\b} \PD{_\d}\TensorField{} \PD{^\d}\TensorField{} -\tfrac{1}{16} \TensorField{}^2 \PD{_\d}\TensorField{} \PD{^\d}\TensorField{} \nonumber\\
    &\quad -2 \TensorField{_\a^\g} \TensorField{^\a^\b} \PD{_\d}\TensorField{_\b^\d} \PD{_\e}\TensorField{_\g^\e} + \TensorField{} \TensorField{^\b^\g} \PD{_\d}\TensorField{_\b^\d} \PD{_\e}\TensorField{_\g^\e} \nonumber\\
    &\quad -4 \TensorField{^\a^\b} \TensorField{^\g^\d} \PD{_\b}\TensorField{_\a_\g} \PD{_\e}\TensorField{_\d^\e} +2 \TensorField{^\a^\b} \TensorField{^\g^\d} \PD{_\g}\TensorField{_\a_\b} \PD{_\e}\TensorField{_\d^\e} \nonumber\\
    &\quad -4 \TensorField{_\a^\g} \TensorField{^\a^\b} \PD{_\g}\TensorField{_\b^\d} \PD{_\e}\TensorField{_\d^\e} +2 \TensorField{} \TensorField{^\b^\g} \PD{_\g}\TensorField{_\b^\d} \PD{_\e}\TensorField{_\d^\e} \nonumber\\
    &\quad +\tfrac{1}{2} \TensorField{_\a_\b} \TensorField{^\a^\b} \PD{_\g}\TensorField{^\g^\d} \PD{_\e}\TensorField{_\d^\e} -\tfrac{1}{4} \TensorField{}^2 \PD{_\g}\TensorField{^\g^\d} \PD{_\e}\TensorField{_\d^\e} \nonumber\\
    &\quad +2 \TensorField{_\a^\g} \TensorField{^\a^\b} \PD{^\d}\TensorField{_\b_\g} \PD{_\e}\TensorField{_\d^\e} - \TensorField{} \TensorField{^\b^\g} \PD{^\d}\TensorField{_\b_\g} \PD{_\e}\TensorField{_\d^\e} \nonumber\\
    &\quad -\tfrac{1}{2} \TensorField{_\a_\b} \TensorField{^\a^\b} \PD{^\d}\TensorField{} \PD{_\e}\TensorField{_\d^\e} +\tfrac{1}{4} \TensorField{}^2 \PD{^\d}\TensorField{} \PD{_\e}\TensorField{_\d^\e} \nonumber\\
    &\quad -2 \TensorField{_\a^\g} \TensorField{^\a^\b} \TensorField{^\d^\e} \PD{_\e}\PD{_\g}\TensorField{_\b_\d} + \TensorField{} \TensorField{^\b^\g} \TensorField{^\d^\e} \PD{_\e}\PD{_\g}\TensorField{_\b_\d} \nonumber\\
    &\quad +2 \TensorField{_\a^\g} \TensorField{^\a^\b} \TensorField{^\d^\e} \PD{_\e}\PD{_\d}\TensorField{_\b_\g} - \TensorField{} \TensorField{^\b^\g} \TensorField{^\d^\e} \PD{_\e}\PD{_\d}\TensorField{_\b_\g} \nonumber\\
    &\quad -2 \TensorField{_\a^\g} \TensorField{^\a^\b} \TensorField{_\b^\d} \PD{_\e}\PD{_\d}\TensorField{_\g^\e} + \TensorField{} \TensorField{_\b^\d} \TensorField{^\b^\g} \PD{_\e}\PD{_\d}\TensorField{_\g^\e} \nonumber\\
    &\quad +\tfrac{1}{2} \TensorField{_\a_\b} \TensorField{^\a^\b} \TensorField{^\g^\d} \PD{_\e}\PD{_\d}\TensorField{_\g^\e} -\tfrac{1}{4} \TensorField{}^2 \TensorField{^\g^\d} \PD{_\e}\PD{_\d}\TensorField{_\g^\e} \nonumber\\
    &\quad +\tfrac{1}{3} \TensorField{_\a^\g} \TensorField{^\a^\b} \TensorField{_\b_\g} \PD{_\e}\PD{_\d}\TensorField{^\d^\e} -\tfrac{1}{4} \TensorField{} \TensorField{_\b_\g} \TensorField{^\b^\g} \PD{_\e}\PD{_\d}\TensorField{^\d^\e} \nonumber\\
    &\quad +\tfrac{1}{24} \TensorField{}^3 \PD{_\e}\PD{_\d}\TensorField{^\d^\e} +2 \TensorField{_\a^\g} \TensorField{^\a^\b} \TensorField{_\b^\d} \PD{_\e}\PD{^\e}\TensorField{_\g_\d} \nonumber\\
    &\quad - \TensorField{} \TensorField{_\b^\d} \TensorField{^\b^\g} \PD{_\e}\PD{^\e}\TensorField{_\g_\d} -\tfrac{1}{2} \TensorField{_\a_\b} \TensorField{^\a^\b} \TensorField{^\g^\d} \PD{_\e}\PD{^\e}\TensorField{_\g_\d} \nonumber\\
    &\quad +\tfrac{1}{4} \TensorField{}^2 \TensorField{^\g^\d} \PD{_\e}\PD{^\e}\TensorField{_\g_\d} -\tfrac{1}{3} \TensorField{_\a^\g} \TensorField{^\a^\b} \TensorField{_\b_\g} \PD{_\e}\PD{^\e}\TensorField{} \nonumber\\
    &\quad +\tfrac{1}{4} \TensorField{} \TensorField{_\b_\g} \TensorField{^\b^\g} \PD{_\e}\PD{^\e}\TensorField{} -\tfrac{1}{24} \TensorField{}^3 \PD{_\e}\PD{^\e}\TensorField{} \nonumber\\
    &\quad +2 \TensorField{^\a^\b} \TensorField{^\g^\d} \PD{_\d}\TensorField{_\g_\e} \PD{^\e}\TensorField{_\a_\b} -\tfrac{1}{2} \TensorField{^\a^\b} \TensorField{^\g^\d} \PD{_\e}\TensorField{_\g_\d} \PD{^\e}\TensorField{_\a_\b} \nonumber\\
    &\quad -2 \TensorField{^\a^\b} \TensorField{^\g^\d} \PD{_\d}\TensorField{_\b_\e} \PD{^\e}\TensorField{_\a_\g} +\tfrac{3}{2} \TensorField{^\a^\b} \TensorField{^\g^\d} \PD{_\e}\TensorField{_\b_\d} \PD{^\e}\TensorField{_\a_\g} \nonumber\\
    &\quad -2 \TensorField{_\a^\g} \TensorField{^\a^\b} \PD{_\g}\TensorField{_\d_\e} \PD{^\e}\TensorField{_\b^\d} + \TensorField{} \TensorField{^\b^\g} \PD{_\g}\TensorField{_\d_\e} \PD{^\e}\TensorField{_\b^\d} \nonumber\\
    &\quad - \TensorField{_\a^\g} \TensorField{^\a^\b} \PD{_\d}\TensorField{_\g_\e} \PD{^\e}\TensorField{_\b^\d} +\tfrac{1}{2} \TensorField{} \TensorField{^\b^\g} \PD{_\d}\TensorField{_\g_\e} \PD{^\e}\TensorField{_\b^\d} \nonumber\\
    &\quad +3 \TensorField{_\a^\g} \TensorField{^\a^\b} \PD{_\e}\TensorField{_\g_\d} \PD{^\e}\TensorField{_\b^\d} -\tfrac{3}{2} \TensorField{} \TensorField{^\b^\g} \PD{_\e}\TensorField{_\g_\d} \PD{^\e}\TensorField{_\b^\d} \nonumber\\
    &\quad +\tfrac{1}{4} \TensorField{_\a_\b} \TensorField{^\a^\b} \PD{_\d}\TensorField{_\g_\e} \PD{^\e}\TensorField{^\g^\d} -\tfrac{1}{8} \TensorField{}^2 \PD{_\d}\TensorField{_\g_\e} \PD{^\e}\TensorField{^\g^\d} \nonumber\\
    &\quad -\tfrac{3}{8} \TensorField{_\a_\b} \TensorField{^\a^\b} \PD{_\e}\TensorField{_\g_\d} \PD{^\e}\TensorField{^\g^\d} \nonumber\\
    &\quad +\tfrac{3}{16} \TensorField{}^2 \PD{_\e}\TensorField{_\g_\d} \PD{^\e}\TensorField{^\g^\d} \bigg]. \label{eq:TensorDerivativeQuartic}
\end{align}
\end{subequations}

\paragraph*{Algebraic interactions} We can also add to~\cref{eq:FP} the cubic and quartic interactions that are purely algebraic. These require the addition of new couplings beyond those of the general tensor theory in~\cref{eq:Tensor}, and we write
\begin{subequations}\label{eq:TensorInteractions}
\begin{align}
	\mathcal{S}(\Coupling{}) &= \frac{1}{\MPl}\int \mathrm{d}^4x\, \bigg[ \Coupling{_7} \TensorField{_\a^\g}\TensorField{^\a^\b}\TensorField{_\b_\g} \nonumber\\
&\hphantom{{}= \frac{1}{\MPl}\int \mathrm{d}^4x\, \bigg[\;}+ \Coupling{_8} \TensorField{}\TensorField{_\a_\b}\TensorField{^\a^\b}
	+ \Coupling{_9} \TensorField{}^3 \bigg], \label{eq:TensorCubic}\\
	\mathcal{S}(\Coupling{}) &= \frac{1}{\MPl^2}\int \mathrm{d}^4x\, \bigg[ \Coupling{_{10}} \TensorField{_\a^\b}\TensorField{_\b^\g}\TensorField{_\g^\d}\TensorField{_\d^\a} \nonumber\\
&\hphantom{{}= \frac{1}{\MPl^2}\int \mathrm{d}^4x\, \bigg[\;}+ \Coupling{_{11}} \TensorField{}\TensorField{_\a^\b}\TensorField{_\b^\g}\TensorField{_\g^\a} \nonumber\\
&\hphantom{{}= \frac{1}{\MPl^2}\int \mathrm{d}^4x\, \bigg[\;}+ \Coupling{_{12}} \TensorField{_\a_\b}\TensorField{^\a^\b}\TensorField{_\g_\d}\TensorField{^\g^\d} \nonumber\\
&\hphantom{{}= \frac{1}{\MPl^2}\int \mathrm{d}^4x\, \bigg[\;}+ \Coupling{_{13}} \TensorField{}^2 \TensorField{_\a_\b}\TensorField{^\a^\b} \nonumber\\
&\hphantom{{}= \frac{1}{\MPl^2}\int \mathrm{d}^4x\, \bigg[\;}+ \Coupling{_{14}} \TensorField{}^4 \bigg]. \label{eq:TensorQuartic}
\end{align}
\end{subequations}
When~\cref{eq:TensorDerivativeInteractions,eq:TensorInteractions} are combined in the sunset and tadpole topologies, the radiative shifts in~\cref{eq:KineticOnlyShifts} are extended to
\begin{subequations}\label{eq:FullShifts}
\begin{align}
    &\delta\Coupling{_1} = \frac{1}{\MPl^2 \pi^2 \varepsilon} \bigg[ \frac{55\,\FPCoupling{_2}^{2}}{36\,\FPCoupling{_1}^{2}} - \frac{65\,\FPCoupling{_2}}{9\,\FPCoupling{_1}^{2}}\,\Coupling{_7} + \frac{20\,\FPCoupling{_2}}{27\,\FPCoupling{_1}^{2}}\,\Coupling{_8} \nonumber\\
    &\quad  + \frac{5}{6\,\FPCoupling{_1}^{2}}\,\Coupling{_7}^{2} + \frac{20}{9\,\FPCoupling{_1}^{2}}\,\Coupling{_7}\Coupling{_8} + \frac{20}{9\,\FPCoupling{_1}^{2}}\,\Coupling{_8}^{2} - \frac{95\,\FPCoupling{_2}}{9\,\FPCoupling{_1}^{2}}\,\Coupling{_{10}} \nonumber\\
    &\quad  - \frac{220\,\FPCoupling{_2}}{9\,\FPCoupling{_1}^{2}}\,\Coupling{_{12}} \bigg] + \dots, \label{eq:FullShift1}
    \\
    &\delta\Coupling{_2} = \frac{1}{\MPl^2 \pi^2 \varepsilon} \bigg[ -\frac{5\,\FPCoupling{_2}^{2}}{72\,\FPCoupling{_1}^{2}} - \frac{35\,\FPCoupling{_2}}{18\,\FPCoupling{_1}^{2}}\,\Coupling{_7} - \frac{140\,\FPCoupling{_2}}{27\,\FPCoupling{_1}^{2}}\,\Coupling{_8} \nonumber\\
    &\quad  + \frac{125}{48\,\FPCoupling{_1}^{2}}\,\Coupling{_7}^{2} + \frac{125}{18\,\FPCoupling{_1}^{2}}\,\Coupling{_7}\Coupling{_8} + \frac{40}{9\,\FPCoupling{_1}^{2}}\,\Coupling{_8}^{2} - \frac{10\,\FPCoupling{_2}}{9\,\FPCoupling{_1}^{2}}\,\Coupling{_{10}} \nonumber\\
    &\quad  - \frac{15\,\FPCoupling{_2}}{2\,\FPCoupling{_1}^{2}}\,\Coupling{_{11}} + \frac{10\,\FPCoupling{_2}}{9\,\FPCoupling{_1}^{2}}\,\Coupling{_{12}} - \frac{10\,\FPCoupling{_2}}{\FPCoupling{_1}^{2}}\,\Coupling{_{13}} \bigg] + \dots, \label{eq:FullShift2}
    \\
    &\delta\Coupling{_3} = \frac{1}{\MPl^2 \pi^2 \varepsilon} \bigg[ \frac{35\,\FPCoupling{_2}}{216\,\FPCoupling{_1}} - \frac{65}{72\,\FPCoupling{_1}}\,\Coupling{_7} - \frac{155}{54\,\FPCoupling{_1}}\,\Coupling{_8} \nonumber\\
    &\quad  + \frac{85}{96\,\FPCoupling{_1}\,\FPCoupling{_2}}\,\Coupling{_7}^{2} + \frac{25}{9\,\FPCoupling{_1}\,\FPCoupling{_2}}\,\Coupling{_7}\Coupling{_8} \nonumber\\
    &\quad  + \frac{25}{9\,\FPCoupling{_1}\,\FPCoupling{_2}}\,\Coupling{_8}^{2} \bigg] + \dots, \label{eq:FullShift3}
    \\
    &\delta\Coupling{_4} = \frac{1}{\MPl^2 \pi^2 \varepsilon} \bigg[ -\frac{5\,\FPCoupling{_2}}{36\,\FPCoupling{_1}} + \frac{55}{18\,\FPCoupling{_1}}\,\Coupling{_7} + \frac{20}{9\,\FPCoupling{_1}}\,\Coupling{_8} \nonumber\\
    &\quad  + \frac{25}{12\,\FPCoupling{_1}\,\FPCoupling{_2}}\,\Coupling{_7}^{2} - \frac{20}{9\,\FPCoupling{_1}\,\FPCoupling{_2}}\,\Coupling{_7}\Coupling{_8} \bigg] + \dots, \label{eq:FullShift4}
    \\
    &\delta\Coupling{_5} = \frac{1}{\MPl^2 \pi^2 \varepsilon} \bigg[ \frac{5\,\FPCoupling{_2}}{27\,\FPCoupling{_1}} - \frac{35}{12\,\FPCoupling{_1}}\,\Coupling{_7} - \frac{25}{54\,\FPCoupling{_1}}\,\Coupling{_8} \nonumber\\
    &\quad  + \frac{15}{8\,\FPCoupling{_1}\,\FPCoupling{_2}}\,\Coupling{_7}^{2} + \frac{40}{9\,\FPCoupling{_1}\,\FPCoupling{_2}}\,\Coupling{_7}\Coupling{_8} \nonumber\\
    &\quad  - \frac{20}{27\,\FPCoupling{_1}\,\FPCoupling{_2}}\,\Coupling{_8}^{2} \bigg] + \dots, \label{eq:FullShift5}
    \\
    &\delta\Coupling{_6} = \frac{1}{\MPl^2 \pi^2 \varepsilon} \bigg[ \frac{55\,\FPCoupling{_2}}{72\,\FPCoupling{_1}} - \frac{65}{18\,\FPCoupling{_1}}\,\Coupling{_7} - \frac{10}{9\,\FPCoupling{_1}}\,\Coupling{_8} \nonumber\\
    &\quad  - \frac{5}{16\,\FPCoupling{_1}\,\FPCoupling{_2}}\,\Coupling{_7}^{2} - \frac{10}{27\,\FPCoupling{_1}\,\FPCoupling{_2}}\,\Coupling{_8}^{2} \bigg] + \dots, \label{eq:FullShift6}
\end{align}
\end{subequations}
where we again use the alternative notation for the RHS free couplings, defined in~\cref{free_redef}. Technical naturalness is manifest in~\cref{eq:FullShifts} if the whole algebraic potential smoothly vanishes with the graviton mass.

\section{Untuned implementation} \label{implementation-appendix}

\paragraph*{Scope} The main ideas from~\cref{implementation} are combined in a prototype system which takes as input some~$\mathcal{S}(\Coupling{})$ corresponding to an untuned theory of the kind considered in~\cref{untuned}. The output is a sample chain corresponding to theories which (i) propagate exclusively one healthy pole in some user-specified~$J^P$ sector, and (ii) comprehensively explore the coupling space. For simplicity,~$\mathcal{S}(\Coupling{})$ is assumed not to contain any gauge symmetries.

\paragraph*{Platforms} An initial application of computer algebra is required to extract the symbolic~$\WaveOperatorJPn{J}{P}{n}$ from~$\mathcal{S}(\Coupling{})$, and this problem was previously solved for any theory of the form given in~\cref{EFTLag} by the \WolframLanguage{} implementation in~\cite{Barker:2024juc,Barker:2025qmw}. The remainder of the pipeline is implemented in~\Python{}, specifically within the \JAX{} framework, which enables just-in-time compilation of all components.\footnote{The use of \JAX{} throughout opens the door, in principle, to an eventual GPU acceleration of the algorithm. Note that, for all the models considered in this work, the sampling procedure has a walltime of minutes on a single CPU core at double precision.}

\paragraph*{Vandermonde inversion} Since~$\WaveOperatorJP{J}{P}$ is Hermitian for real~$k$, its determinant is a polynomial in even powers of~$k$ and so admits the expansion
\begin{equation}\label{eq:DetExpansion}
\det\WaveOperatorJP{J}{P} = \sum_{j=0}^{n_{J^P}}\, \DetCoeff{j}{J}{P}\, z^j,\qquad z \equiv k^2.
\end{equation}
The symbolic computation of~\cref{eq:DetExpansion} in the~$\Coupling{}$ is to be avoided, since it may contain many thousands of monomials. At each sampled~$\Coupling{}$, the determinant is thus evaluated numerically at~$n_{J^P}+1$ `test' momenta given by
\begin{equation}\label{eq:DetNodes}
\VanZ{i} = i,\quad \VanK{i} = \sqrt{\VanZ{i}},\qquad i = 0,1,\ldots,n_{J^P}.
\end{equation}
The coefficients~$\DetCoeff{j}{J}{P}$ in~\cref{eq:DetExpansion} are then recovered by inverting the Vandermonde system
\begin{equation}\label{eq:Vandermonde}
d_i \equiv \sum_{j=0}^{n_{J^P}}\, \DetCoeff{j}{J}{P}\, \VanZ{i}^{j} \;=\; \sum_j \VanV{i}{j}\, \DetCoeff{j}{J}{P},
\end{equation}
where~$\VanV{i}{j} \equiv i^{j}$. This matrix depends only on~$n_{J^P}$, not on~$\Coupling{}$, so its inverse can be precomputed.

\paragraph*{Sampling} To make sure that only the target pole is propagating, it suffices to send all other poles to infinity. The residual vector~$\mathsf{r}(\theta)$ is constructed by stacking all the~$\DetCoeff{j}{J}{P}$ which are required to vanish in order to produce this decoupling. This means that the index~$j$ ranges over all~$j\geq 1$ for non-target~$J^P$ and over~$j\geq 2$ for the target sector. Nested sampling is then performed, with the likelihood
\begin{equation}\label{eq:Likelihood_Det}
\log\Likelihood = -\lvert\mathsf{r}(\theta)\rvert^{2}.
\end{equation}
Note that~\cref{eq:Likelihood_Det} is actually somewhat simpler than our proposal for the tuned case in~\cref{eq:Likelihood}.

We use the \BlackJAX{} nested sampler~\cite{cabezas2024blackjax,yallup2026nested} with~$10^4$ live points and~$5(N-1)$ inner steps, deleting~$10^3$ points at each iteration, and with a log-likelihood threshold of~$-10^{-10}$. The prior is uniform on the hypersphere, and a geodesic (i.e., great circle) stepper function respects this natural measure.

As mentioned in~\cref{implementation}, nested sampling is not inherently good at root-finding. Rather, it efficiently samples distributions in high-$N$ parameter spaces. Here, nested sampling is used to rapidly produce an initial pool of samples whose volume is distributed around small values of~$\mathsf{r}(\theta)$. It is expected that a large fraction of this pool will be sufficiently optimised for subsequent refinement, by methods which scale less well with~$N$.

\paragraph*{Refining} For each of the samples in the resulting chain, the residual vector~$\mathsf{r}(\theta)$ is further reduced through Levenberg--Marquardt (LM) optimisation. This procedure requires the Jacobian
\begin{equation}\label{eq:LMJacobian}
\mathsf{J}(\theta) \equiv \frac{\partial\mathsf{r}(\theta)}{\partial\theta},
\end{equation}
which can be obtained by forward-mode automatic differentiation through~\cref{eq:DetExpansion,eq:DetNodes,eq:Vandermonde}.

We use the \Optimistix{} LM implementation~\cite{optimistix2024}, with up to~$30$ iterations, and relative and absolute tolerances of~$10^{-12}$. Samples for which the LM process fails are dropped from the chain, along with samples for which the process yields a final~$\lvert\mathsf{r}(\theta)\rvert>10^{-6}$.

\paragraph*{Measuring} By construction, the refined samples are known to propagate one (relatively) light pole, with all other poles separated by a substantial mass hierarchy. The masses themselves have not yet been computed, however. This is done directly by the Frobenius companion matrix method, since the~$\DetCoeff{i}{J}{P}$ themselves are by this point known numerically. The residues are then computed from~\cref{eq:NoGhost}.

In summary, the sampling and refining procedures identify precise points on the target hypersurface, whilst measuring determines the hypersurface boundaries.

\paragraph*{Reweighting}  The distribution of the surviving points on the hypersurface is affected by~$\mathsf{r}(\theta)$ and its gradient through~\cref{eq:Likelihood_Det,eq:LMJacobian}, and this is an unavoidable consequence of sampling and root-finding. Since the points themselves are fundamentally produced at cost in our approach, the `cheapest' procedure is to reweight them in the final chain. A~$k$-nearest-neighbour ($k$NN) density estimate is used to restore uniformity with respect to the original prior on the hypersphere; the details are provided in~\cref{local_dimensionality}. Further details are available at~\cite{supp_materials}.

\bibliography{Manuscript}

\end{document}